\documentclass[12pt]{article}
\usepackage{amssymb}
\renewcommand{\theequation}{\arabic{section}.\arabic{equation}}

\begin{document}

%************************** Text Begins here ******************************

%  Greek letters

\def\a{\alpha}
\def\b{\beta}
\def\d{\delta}
\def\e{\epsilon}
\def\g{\gamma}
\def\h{\mathfrak{h}}
\def\k{\kappa}
\def\l{\lambda}
\def\o{\omega}
\def\p{\wp}
\def\r{\rho}
\def\t{\tau}
\def\s{\sigma}
\def\z{\zeta}
\def\x{\xi}
 \def\A{{\cal{A}}}
 \def\B{{\cal{B}}}
 \def\C{{\cal{C}}}
 \def\D{{\cal{D}}}
\def\P{{\cal{P}}}
\def\G{\Gamma}
\def\K{{\cal{K}}}
\def\O{\Omega}
\def\R{\bar{R}}
\def\S{{\cal{S}}}
\def\T{{\cal{T}}}
\def\L{\Lambda}
\def\GL{U_q(gl(m|n))}
\def\gl{U_q(gl(2|1))}
\def\Zb{\mathbb{Z}}
\def\Cb{\mathbb{C}}

\def\R{\overline{R}}
% Shorthands for \begin{equation} and the like

\def\beq{\begin{equation}}
\def\eeq{\end{equation}}
\def\bea{\begin{eqnarray}}
\def\eea{\end{eqnarray}}
\def\ba{\begin{array}}
\def\ea{\end{array}}
\def\no{\nonumber}
\def\le{\langle}
\def\re{\rangle}
\def\lt{\left}
\def\rt{\right}

\newtheorem{Theorem}{Theorem}
\newtheorem{Definition}{Definition}
\newtheorem{Proposition}{Proposition}
\newtheorem{Lemma}{Lemma}
\newtheorem{Corollary}{Corollary}
\newcommand{\proof}[1]{{\bf Proof. }
        #1\begin{flushright}$\Box$\end{flushright}}

\baselineskip=20pt

%%%%%%%%%%%%%%%%%%%%%%%%%%%%%%%%%%%%%%%%%%%%%%%%%%%%%%%%%%%%
%                                                          %
%  Title page                                              %
%                                                          %
%%%%%%%%%%%%%%%%%%%%%%%%%%%%%%%%%%%%%%%%%%%%%%%%%%%%%%%%%%%%
\newfont{\elevenmib}{cmmib10 scaled\magstep1}
\newcommand{\preprint}{
   %\begin{flushleft}
    % \elevenmib Yukawa\, Institute\, Kyoto\\
   %\end{flushleft}\vspace{-1.3cm}
   \begin{flushright}\normalsize
     {\tt hep-th/0503003} \\ March 2005
   \end{flushright}}
\newcommand{\Title}[1]{{\baselineskip=26pt
   \begin{center} \Large \bf #1 \\ \ \\ \end{center}}}
\newcommand{\Author}{\begin{center}
   \large \bf
Wen-Li Yang$,{}^{a,b}$
 ~Yao-Zhong Zhang${}^b$~and Shao-You Zhao$^{b,c}$ \footnote{Corresponding author: syz@maths.uq.edu.au}  \end{center}}
\newcommand{\Address}{\begin{center}

     ${}^a$ Institute of Modern Physics, Northwest University,
     Xian 710069, P.R. China\\
     ${}^b$ Department of Mathematics, University of Queensland, Brisbane, QLD 4072,
     Australia\\
     ${}^c$ Department of Physics, Beijing Institute of
     Technology, Beijing 100081, China
   \end{center}}
\newcommand{\Accepted}[1]{\begin{center}
   {\large \sf #1}\\ \vspace{1mm}{\small \sf Accepted for Publication}
   \end{center}}
\preprint
\thispagestyle{empty}
\bigskip\bigskip\bigskip

\Title{Drinfeld twists and algebraic Bethe ansatz of the
supersymmetric model associated with $\GL$ } \Author

\Address
\vspace{1cm}

\begin{abstract}
We construct the Drinfeld twists (or factorizing $F$-matrices) of
the  supersymmetric  model associated with quantum superalgebra
$\GL$, and obtain the completely symmetric representations of the
creation operators of the model in the $F$-basis provided by the
$F$-matrix.  As an application of our general results, we present
the explicit expressions of the Bethe vectors in the  $F$-basis
for the $\gl$-model (the quantum $t$-$J$ model).

\vspace{1truecm}
%\noindent {\it PACS:} 03.65.Fd; 04.20.Jb;
%05.30.-d; 75.10.Jm

\end{abstract}
\newpage
%%%%%%%%%%%%%%%%%%%%%%%%%%%%%%%%%%%%%%%%%%%%%%%%%%%%%%%%%%%%%%%
%                                                             %
%  1. Introduction                                            %
%                                                             %
%%%%%%%%%%%%%%%%%%%%%%%%%%%%%%%%%%%%%%%%%%%%%%%%%%%%%%%%%%%%%%%
\section{Introduction}
\label{intro} \setcounter{equation}{0}

It was realized in \cite{Maillet96} that for the XXX or XXZ spin
chain systems, there exists a non-degenerate lower-triangular
$F$-matrix (the Drinfeld twists) \cite{Drinfeld83} in terms of
which the $R$-matrix of the system is factorized:
\begin{equation}
R_{12}(u_1,u_2)=F^{-1}_{21}(u_2,u_1)F_{12}(u_1,u_2),
\label{eq:R-FF}
\end{equation}
where the $R$-matrix acts on the tensor space $V\otimes V$ with
$V$ being a 2-dimensional $U_q(gl(2))$-module. In the basis
provided by the $N$-site $F$-matrix, i.e. the so-call $F$-basis,
the entries of the monodromy matrices of the models appear in
completely symmetric forms. As a result the Bethe vectors of the
models are dramatically simplified and can be written down
explicitly. These results enabled the authors in
\cite{Kitanine98,Izergin98} to compute form factors, correlation
functions \cite{Kor93} and spontaneous magnetizations of the
systems analytically and explicitly.

The results of \cite{Maillet96} were generalized to  other
systems. In \cite{Terras99}, the Drinfeld twists associated with
any finite-dimensional irreducible representations of Yangian
$Y[gl(2)]$ were investigated. In \cite{Albert00}, Albert et al
constructed the $F$-matrix of the rational $gl(m)$ Heisenberg
model, obtained a polarization free representation of its creation
operators and resolved the hierarchy of its nested Bethe vectors.
In \cite{Albert0005,Albert0007}, the Drinfeld twists of the
elliptic XYZ and Belavin models were constructed. Recently we have
successfully constructed the Drinfeld twists for the rational
$gl(m|n)$ supersymmetric model and resolved the hierarchy of its
nested Bethe vectors in the $F$-basis \cite{zsy04,ZYZ05}. Quantum
integrable models associated with Lie superalgebras
\cite{Per81,Kul82,Kul86} are physically important because they
give strongly correlated fermion models of superconductivity (e.g.
\cite{Ess92,Bra95}).

In this paper, we extend our results in \cite{zsy04,ZYZ05} to the
quantum (or q-deformed)  supersymmetric  model associated the
quantum superalgebra $\GL$ (including  quantum supersymmetric
$t$-$J$ model as a special case). Such a generalization is {\it
non-trivial} due to the following fact. It is well-known that the
$gl(m|n)$  rational model  has  $gl(m|n)$ symmetry which enables
one to express the creation operators $C_{i}(u)$ in terms of the
element $T_{m+n,m+n}(u)$ of the monodromy matrix $T(u)$ and the
generators of $gl(m|n)$ by (anti)commutation relations
\cite{Albert00,ZYZ05}. However, the corresponding quantum model is
not $\GL$ invariant (unless appropriate boundary conditions are
imposed). One of the consequences is that  the creation operators
$C_i(u)$ of the quantum model {\it cannot\/} be expressed in terms
of $T_{m+n,m+n}(u)$ and the generators of $\GL$  by {\it simple\/}
q-(anti)commutation relations. Indeed, it is found in this paper
that {\it extra quantum correction terms} are needed, due to the
non-trivial coproduct structure of the quantum superalgebra.
Having  found such a new recursive relation (\ref{Recursive}) and
constructed the factorizing $F$-matrices of the quantum model, we
obtain the symmetric representations of the creation operators of
the monodromy matrix in the $F$-basis. These results make possible
a complete resolution of  the hierarchy of the nested Bethe
vectors of the $\GL$ model. As an example, we give the explicit
expressions of the Bethe vectors of the quantum $t-J$ model
associated with $\gl$.

The present paper is organized as follows. In section 2, we
introduce some basic notation on the quantum superalgebra
$U_q(gl(m|n))$. In section 3, we derive the recursive relation
between the  elements of the monodromy matrix and the generators
of $\GL$. In section 4, we construct the $F$-matrix and its
inverse of  the $\GL$ model. In section 5, we obtain the symmetric
representations of the creation operators in the $F$-basis. As an
application of our general results, the hierarchy of the nested
Bethe vectors of the $\gl$ model is resolved in section 6. We
conclude the paper by offering some discussions in section 7. Some
detailed technical derivations are given in Appendices A-B.

%%%%%%%%%%%%%%%%%%%%%%%%%%%%%%%%%%%%%%%%%%%%%%%%%%%%%%%%%%%%%%%
%                                                             %
%  2. Quantum superalgebra $U_q(gl(m|n))$                     %
%                                                             %
%                                                             %
%                                                             %
%%%%%%%%%%%%%%%%%%%%%%%%%%%%%%%%%%%%%%%%%%%%%%%%%%%%%%%%%%%%%%%

\section{Quantum superalgebra $\GL$}
\label{QSA} \setcounter{equation}{0}

Let us fix two non-negative integers $n$, $m$ such that $n+m\geq
2$  and a positive integer $N\,(\geq 2)$, and a generic complex
number $\eta$ such that the q-deformation parameter, which is
defined by $q=e^{\eta}$, is not a root of unity. Let $V$ be a
$\Zb_2$-graded $(n+m)$-dimensional vector space with the
orthonormal basis $\{|i\rangle,\,i=1,\ldots,n+m\}$. The
$\Zb_2$-grading is chosen as:
$[1]=\ldots=[m]=1,\,[m+1]=\ldots=[m+n]=0$.

\vskip0.2in

\begin{Definition} The quantum superalgebra $\GL$ is a $\Zb_2$-graded
unital associative superalgebra   generated by the generators
$E^{i,i},\,(i=1,\ldots,n+m)$ and
$E^{j,j+1},\,E^{j+1,j}\,(j=1,\ldots,n+m-1)$  with the
$\Zb_2$-grading $[E^{i,i}]=0$, $[E^{j+1,j}]=[E^{j,j+1}]=[j]+[j+1]$
by the relations: \bea &&[E^{i,i},E^{i',i'}]=0,\, [E^{i,i},
E^{j,j+1}]=(\d_{i,j}-\d_{i,j+1})E^{i,j+
1},~i'=1,\ldots,n+m,\label{Def-1}\\
&& [E^{i,i}, E^{j+1,j}]=(\d_{i,j+1}-\d_{i,j})E^{j+1,j},\label{Def-2}\\
&&[E^{j,j+1},E^{j'+1,j'}]=(-1)^{[j]}\d_{j,j'}\frac{q^{h^j}-q^{-h^j}}{q-q^{-1}},
~j'=1,\ldots,n+m-1,\label{Def-3}\eea and the Serre relations: \bea
&&(E^{m,m+1})^2=(E^{m+1,m})^2=0,\no\\
&&[E^{j,j+1},E^{j',j'+1}]=[E^{j+1,j},E^{j'+1,j'}]=0,~|j-j'|\geq
2,\label{Serre}\\
&&(E^{j,j+1})^2E^{j\pm 1,j\pm 1+1}-(q+q^{-1})E^{j,j+1}E^{j\pm
1,j\pm 1+1}E^{j,j+1}\no\\
&&\quad\quad\quad\quad+E^{j\pm 1,j\pm 1+1}(E^{j,j+1})^2=0,~j\neq
m,\no\\
&&(E^{j+1,j})^2E^{j\pm 1+1,j\pm 1}-(q+q^{-1})E^{j+1,j}E^{j\pm
1+1,j\pm 1}E^{j+1,j}\no\\
&&\quad\quad\quad\quad+E^{j\pm 1+1,j\pm 1}(E^{j+1,j})^2=0,~j\neq
m,\no\eea where $h^{j}=(-1)^{[j]}E^{j,j}-(-1)^{[j+1]}E^{j+1,j+1}$.
In addition to the above Serre relations, there exist also extra
Serre relations \cite{Yam96} which we omit.
\end{Definition}

\vskip 0.2in

\noindent Here and throughout, we adopt the convention:\bea
[x,y]=xy-(-1)^{[x][y]}yx, ~~x,y\in \GL.\no\eea One can easily see
that the $\Zb_2$-graded vector space $V$ supplies  the fundamental
$\GL$-module
 and the generators of $\GL$ are represented in this space by
\bea \pi(E^{i,i})=e_{i,i},~
\pi(E^{j,j+1})=e_{j,j+1},~\pi(E^{j+1,j})=e_{j+1,j},\label{Fund}\eea
where $e_{i,j}\in{\rm End}(V)$ is the elementary matrix with
elements $(e_{i,j})^l_k=\d_{jk}\d_{il}$.

$\GL$ is a $\Zb_2$-graded triangular Hopf superalgebra endowed
with $\Zb_2$-graded algebra  homomorphisms that are coproduct
$\Delta$: $\GL\longrightarrow \GL\otimes\GL$ defined by \bea
&&\Delta(E^{i,i})=1\otimes E^{i,i}+E^{i,i}\otimes
1,~i=1,\ldots,n+m,\label{copr-1}\\
&&\Delta(E^{j,j+1})=1\otimes E^{j,j+1}+E^{j,j+1}\otimes q^{h^j},\label{copr-2}\\
&&\Delta(E^{j+1,j})=q^{-h^{j}}\otimes E^{j+1,j}+E^{j+1,j}\otimes
1,\label{copr-3}\eea and counit $\e$: $\GL\longrightarrow \Cb$
defined by \bea
\e(E^{j,j+1})=\e(E^{j+1,j})=\e(E^{i,i})=0,~~\e(1)=1,\no\eea and a
$\Zb_2$-graded algebra antiautomorphism (antipode) $S$:
$\GL\longrightarrow \GL$ given by \bea
S(E^{j,j+1})=-E^{j,j+1}q^{-h^j},~
S(E^{j+1,j})=-q^{h^j}E^{j+1,j},~S(E^{i,i})=-E^{i,i}. \no\eea
Multiplications of tensor products are $\Zb_2$ graded: \bea
(x\otimes y)(x'\otimes y')=(-1)^{[y][x']}xx'\otimes yy',\no \eea
for homogeneous elements $x,y,x',y'\in \GL$ and where $[x]\in
\Zb_2$ denotes the grading of $x$. It should be pointed out that
the antipode satisfies the following equation, for homogeneous
elements $x,y\in \GL$, \bea S(xy)=(-1)^{[x][y]} S(y)S(x),\no \eea
and generalizes to inhomogeneous elements through linearity. The
coproduct, counit and antipode satisfy the following relations,
$~~\forall x\in\GL$ : \bea &&(\Delta \otimes {\rm
id})\Delta(x)=({\rm id}\otimes
\Delta)\Delta (x), \label{Hopf}\\
&&(\e\otimes {\rm id})\Delta(x)=x=({\rm id}\otimes\e)\Delta(x),\no\\
&&m(S\otimes {\rm id})\Delta(x)=m({\rm id}\otimes
S)\Delta(x)=\e(x),\no\eea where $m$ denote the product of any two
elements of $\GL$, i.e., $m(x\otimes y)=xy$ for $x,y\in \GL$.

The generators $\{E^{j,j+1}\}$ ($\,\{E^{j+1,j}\}$) are the simple
raising (lowering) generators of $\GL$ associated with the simple
roots. Thanks to the Serre relations, the other generators
associated with the non-simple roots (called the non-simple
generators) can be uniquely constructed through the simple ones by
the following relations:\bea &&E^{\a,\g}=E^{\a,\b}E^{\b,\g}
   -q^{-(-1)^{[\b]}}E^{\b,\g}E^{\a,\b},~~1\leq\a<\b<\g\leq n+m,
   \label{non-simple1}\\
   &&E^{\g,\a}=E^{\g,\b}E^{\b,\a}
   -q^{(-1)^{[\b]}}E^{\b,\a}E^{\g,\b},~~1\leq\a<\b<\g\leq n+m.
   \label{non-simple2}\eea
The coproduct, counit and antipode of the non-simple generators
can be obtained through those of the simple ones. Here, we give
the coproduct of non-simple generators which will be used later.

\vskip0.1in

\begin{Lemma}
The coproduct of the non-simple generators is \bea
&&\Delta(E^{\g,\g-l})=q^{-\sum_{k=1}^{l}h^{\g-k}}\otimes
E^{\g,\g-l}+E^{\g,\g-l}\otimes 1\no\\
&&\quad+\hspace{-0.1truecm}\sum_{i=1}^{l-1}
(1\hspace{-0.1truecm}-\hspace{-0.1truecm} q^{2(-1)^{[\g-l+i]}})
q^{-\sum_{k=1}^{l-i}h^{\g-k}}E^{\g-l+i,\g-l}\hspace{-0.1truecm}
\otimes\hspace{-0.1truecm}
E^{\g,\g-l+i},~\g-l\geq 1\,{\rm and}\,l\geq 2,\\
&&\Delta(E^{\g,\g+l})=1\otimes E^{\g,\g+l}+E^{\g,\g+l}\otimes
q^{\sum_{k=0}^{l-1}h^{\g+k}}\no\\
&&\quad+\sum_{i=1}^{l-1}(1\hspace{-0.1truecm}-\hspace{-0.1cm}q^{-2(-1)^{[\g+i]}})
E^{\g+i,\g+l}\hspace{-0.truecm}\otimes\hspace{-0.truecm}
E^{\g,\g+i}q^{\sum_{k=i}^{l-1}h^{\g+k}}, ~\g+l\leq n+m\,{\rm
and}\,l\geq 2. \eea
\end{Lemma}

\vskip0.1in

\noindent {\it Proof}. This lemma can be proved by induction using
the coproducts of the simple generators
(\ref{copr-1})-(\ref{copr-3}), the definitions of the non-simple
generators (\ref{non-simple1})-(\ref{non-simple2}) and the fact
that the coproduct is an algebra homomorphism, as well as
(\ref{Def-1})-(\ref{Def-3}) and the Serre relation
(\ref{Serre}).~~ $\Box$

%%%%%%%%%%%%%%%%%%%%%%%%%%%%%%%%%%%%%%%%%%%%%%%%%%%%%%%%%%%%%%%%%%%
%                                                                 %
%     Recursive relation between monodromy matrix elements        %
%                     and $\GL$ generators                        %
%                                                                 %
%                                                                 %
%%%%%%%%%%%%%%%%%%%%%%%%%%%%%%%%%%%%%%%%%%%%%%%%%%%%%%%%%%%%%%%%%%%

\section{Recursive relation between monodromy matrix elements and $\GL$ generators}
\label{RM} \setcounter{equation}{0}

Let $R\in {\rm End}(V\otimes V)$ be the $R$-matrix associated with
the fundamental  $U_q(gl(m|n))$-module $V$. The $R$-matrix depends
on the difference of two spectral parameters $u_1$ and $u_2$
associated with two copies of $V$, and is, in the present grading,
given by \cite{Kul82,Kul86,Baz88}
\begin{eqnarray}
 R_{12}(u_1,u_2)&=&R_{12}(u_1-u_2)\nonumber\\
 &=&c_{12}\sum_{i=1}^m e_{i,i}\otimes e_{i,i}
  +\sum_{i=m+1}^{m+n} e_{i,i}\otimes e_{i,i}
  +a_{12}\sum_{i\ne j=1}^{m+n} e_{i,i}\otimes e_{j,j}\nonumber\\ &&
  \mbox{}
  +b^-_{12}\sum_{i>j=1}^{m+n}(-1)^{[j]}e_{i,j}\otimes e_{j,i}
  +b^+_{12}\sum_{j>i=1}^{m+n}(-1)^{[j]}e_{i,j}\otimes
  e_{j,i},  %\nonumber\\
\end{eqnarray}
where
\begin{eqnarray}
&& a_{12}=a(u_1,u_2)\equiv {\sinh(u_1-u_2)\over
             \sinh(u_1-u_2+\eta)},\quad \quad
   b_{12}^\pm=b^\pm(u_1,u_2)\equiv{e^{\pm(u_1-u_2)}\sinh\eta\over
          \sinh(u_1-u_2+\eta)},\quad\quad\\
&& c_{12}=c(u_1,u_2)\equiv{\sinh(u_1-u_2-\eta)\over
\sinh(u_1-u_2+\eta)},
\end{eqnarray}
and $\eta$ is the so-called crossing parameter. One can easily
check that the $R$-matrix satisfies the unitary relation
\begin{equation}
R_{21}R_{12}=1. \label{eq:unitary}
\end{equation}
Let us introduce the $(N+1)$-fold tensor product space
$\stackrel{N+1}{\overbrace{V\otimes V\cdots\otimes V}}$, whose
components are labelled by $0,1\ldots,N$ from the left to the
right. As usual, the $0$-th space, denoted by $V_0$ ( $V_i$ for
the $i$-th space), corresponds to the auxiliary space and the
other $N$ spaces constitute the quantum space
${\cal{H}}=\stackrel{N}{\overbrace{V\otimes V\cdots\otimes V}}$.
Moreover, for each factor space $V_i$, $i=0,\ldots,N$, we
associate a complex parameter $z_i$. The parameter associated with
the $0$-th space is usually called the {\it spectral} parameter
which is set  to $z_0=u$ in this paper, and the other parameters
are called the {\it inhomogeneous} parameters. In this paper we
always assume that all the complex parameters $u$ and
$\{z_i|i=1,\ldots,N\}$ be {\it generic} ones. Hereafter  we adopt
the standard notation: for any matrix $A\in {\rm End}(V)$,  $A_j$
(or $A_{(j)}$) is an embedding operator in the tensor product
space, which acts as $A$ on the $j$-th space and as an identity on
the other factor spaces; $R_{ij}=R_{ij}(z_i,z_j)$ is an embedding
operator of R-matrix in the tensor product space, which acts as an
identity on the factor spaces except for the $i$-th and $j$-th
ones.

The $R$-matrix satisfies the graded Yang-Baxter equation (GYBE)
\begin{equation}
R_{12}R_{13}R_{23}=R_{23}R_{13}R_{12}.\label{GYBE}
\end{equation}
In terms of the matrix elements defined by \bea
R(u)(|i'\rangle\otimes
|j'\rangle)=\sum_{i,j}R(u)^{i'j'}_{ij}(|i\rangle\otimes
|j\rangle), \no\eea the GYBE reads \bea &&
\sum_{i',j',k'}R(u_1-u_2)^{i'j'}_{ij}R(u_1-u_3)^{i''k'}_{i'k}R(u_2-u_3)^{j''k''}_{j'k'}
    (-1)^{[j']([i']+[i''])}\no\\
&=&\sum_{i',j',k'}R(u_2-u_3)^{j'k'}_{jk}R(u_1-u_3)^{i'k''}_{ik'}R(u_1-u_2)^{i''j''}_{i'j'}
    (-1)^{[j']([i]+[i'])}.\no
\eea Besides the GYBE, the $R$-matrix satisfies the following
relation \bea
&&R_{12}\Delta(x)=\P_{12}\Delta(x)\P^{-1}_{12}R_{12},\label{Bas-1}\eea
where $\P_{12}$ is the superpermutation operator, i.e.,
$\P_{12}(|i\rangle\otimes|j\rangle=(-1)^{[i][j]}|j\rangle\otimes|i\rangle$.

Using the coproduct structure of $\GL$, one can define the action
of $\GL$ on the $(N+1)$-fold tensor product space. For any $ x\in
\GL$, let us denote the action of $x$ on the $(N+1)$-fold tensor
product space by $(x)_{0\ldots N}$:
 \bea
(x)_{0\ldots N}=\Delta^{(N)}(x)=({\rm id}\otimes
\Delta^{(N-1)})\Delta(x). \eea By a straightforward calculation,
one has

\vskip0.1in

\begin{Lemma}\bea
(E^{i,i})_{0\ldots N}&=&\sum_{k=0}^{N}E^{i,i}_{(k)},~i=1,\ldots,n+m,\\
(E^{j,j+1})_{0\ldots
N}&=&\sum_{k=0}^NE^{j,j+1}_{(k)}q^{\sum_{i=k+1}^Nh^j_{(i)}},
~j=1,\ldots,n+m-1,\\
(E^{j+1,j})_{0\ldots
N}&=&\sum_{k=0}^Nq^{-\sum_{i=0}^{k-1}h^j_{(i)}}E^{j+1,j}_{(k)},
~j=1,\ldots,n+m-1,\eea where $E^{i,j}_{(k)}$ is the embedding of
$e_{i,j}$ in the tensor product space, which acts as $e_{i,j}$ on
the $k$-th space and as identity on the other factor spaces.
\end{Lemma}

\vskip0.1in

\noindent The actions of non-simple generators can be obtained
from those of simple ones through (\ref{non-simple1}) and
(\ref{non-simple2}).

Let $\S_{N+1}$ denote the  permutation group of the $N+1$ space
labels  $(0,\ldots,N)$. The GYBE (\ref{GYBE}) and unitary relation
(\ref{eq:unitary}) of $R$-matrix allow one to introduce the
following mapping.

\vskip0.in
\begin{Definition} One can define a mapping from $\S_{N+1}$ to ${\rm
End}(V_0\otimes{\cal{H}})$ which associate in a unique way an
element  $R^{\s}_{0\ldots N}\in {\rm End}(V_0\otimes {\cal{H}})$
to any element $\s$ of the permutation group $\S_{N+1}$. The
mapping has the following composition law \bea R^{\s\s'}_{0\ldots
N}=\P^{\s}\,R^{\s'}_{0\ldots N}\,(\P^{\s})^{-1} \,R^{\s}_{0\ldots
N}=R^{\s'}_{\s(0\ldots N)}\,R^{\s}_{0\ldots N},~\forall \s,\s'\in
\S_{N+1},\label{elemet-1}\eea where $\P^{\s}$ is the
$\Zb_2$-graded permutation operator on the tensor product space,
i.e., %$\quad\quad$
$\P^{\s}|i_0\rangle_{(0)}\ldots|i_N\rangle_{(N)}
=|i_0\rangle_{(\s(0))}\ldots|i_N\rangle_{(\s(N))}$. For any
elementary permutation $\s_{j}$ with
$\s_j(0,\ldots,j,j+1,\ldots,N)=(0,\ldots,j+1,j,\ldots,N)$,
$j=0,\ldots,N$,  the corresponding $R^{\s_j}_{0\ldots N}$ is \bea
R^{\s_j}_{0\ldots N}=R_{j\,j+1}.\label{elemet}\eea
\end{Definition}
\vskip0.1in

\noindent For any element $\s\in \S_{N+1}$, the corresponding
$R^{\s}_{0\ldots N}$ can be constructed through (\ref{elemet-1})
and (\ref{elemet}) as follows. Let $\s$ be decomposed in a minimal
way in terms of elementary permutation as
$\s=\s_{\b_1}\ldots\s_{\b_p}$ where the positive integer $p$ is
the length of $\s$. The composition law enables one to obtain  the
expression of the associated $R^{\s}_{0\ldots N}$. The GYBE
(\ref{GYBE}) and (\ref{eq:unitary}) guarantee the uniqueness of
$R^{\s}_{0\ldots N}$. For the  special element $\s_c$ of
$\S_{N+1}$, \bea \s_c=\s_0\s_1\ldots\s_{N-1},~~{\rm
namely},~~\s_c(0,1,\ldots,N)=(1,2,\ldots,N,0), \label{Sc}\eea the
associated $R^{\s_c}_{0\ldots N}$ is given by \bea T(u)\equiv
T_0(u)=T_{0,1\ldots N}(u)= R^{\s_c}_{0\ldots
N}=R_{0\,N}R_{0\,N-1}\ldots\,R_{01}.\label{Mon}\eea Thus
$R^{\s_c}_{0\ldots N}$ is the quantum monodromy matrix $T(u)$ of
the $\GL$ spin chain on an $N$-site lattice. By the GYBE, one may
prove that the monodromy matrix satisfies the GYBE \bea
R_{00'}(u-v)T_0(u)T_{0'}(v)=T_{0'}(v)T_{0}(u)R_{00'}(u-v).\label{eq:GYBE-1}
\eea Define the transfer matrix $t(u)$
\begin{eqnarray}
t(u)=str_0T(u),\label{de:t}
\end{eqnarray}
where $str_0$ denotes the supertrace over the auxiliary space.
Then the Hamiltonian of our model is given by
\begin{equation}
H={d\ln t(u)\over du}|_{u=0}. \label{de:H}
\end{equation}
This model is integrable thanks to the commutativity of the
transfer matrix for different parameters,
\begin{equation}
[t(u),t(v)]=0,
\end{equation}
which can be verified by using the GYBE.

The fundamental relation (\ref{Bas-1}) and the co-associativity
(\ref{Hopf}) of the coproduct of $\GL$ enable one to prove the
following result, using the procedure  similar to that in
\cite{Maillet96} for the non-super case, \vskip0.1in
\begin{Proposition} The mapping defined in Definition 2 satisfies the
following relation
 \bea R^{\s}_{0\ldots N}\,\,(x)_{0\ldots N}=
\P^{\s}\,\,(x)_{0\ldots N}\,\,(\P^{\s})^{-1}\,R^{\s}_{0\ldots N},
~\forall x\in \GL,~\s\in\S_{N+1}.\label{F-def}\eea
\end{Proposition}
\vskip0.1in

One may decompose the monodromy matrix $T(u)$ in terms of the
basis of ${\rm End}(V_0)$ as\bea
T(u)=\sum_{i,j=1}^{n+m}T_{i,j}(u)E^{i,j}_{(0)}\equiv
\sum_{i,j=1}^{n+m} T_{i,j}(u)e_{i,j}, \label{def-T}\eea where the
matrix elements $T_{i,j}(u)$ are operators acting on the quantum
space ${\cal{H}}$ and have  the $\Zb_2$-grading:
$[T_{i,j}(u)]=[e_{i,j}]=[i]+[j]$. Similarly, for the quantities
defined in Lemma 2, we have the decomposition: \bea
(E^{i,i})_{0\ldots N}&=&E^{i,i}_{(0)}+\sum_{k=1}^{N}E^{i,i}_{(k)}
=e_{i,i}+(E^{i,i})_{1\ldots N},~i=1,\ldots,n+m,\label{t-eq-1}\\
(E^{j,j+1})_{0\ldots
N}&=&E^{j,j+1}_{(0)}\,q^{\sum_{k=1}^{N}h^{j}_{(k)}}
+\sum_{k=1}^NE^{j,j+1}_{(k)}q^{\sum_{i=k+1}^Nh^j_{(i)}}\no\\
&=&e_{j,j+1}\,q^{(h^j)_{1\ldots N}}+(E^{j,j+1})_{1\ldots N},~j=1,\ldots,n+m-1,\\
(E^{j+1,j})_{0\ldots N}&=&E^{j+1,j}_{(0)}+q^{-h^j_{(0)}}
\sum_{k=1}^Nq^{-\sum_{i=1}^{k-1}h^j_{(i)}}E^{j+1,j}_{(k)}\no\\
&=&e_{j+1,j}+q^{-h_j}\,(E^{j+1,j})_{1\ldots N},
~j=1,\ldots,n+m-1,\label{t-eq-2}\eea where
$h_j=(-1)^{[j]}e_{j,j}-(-1)^{[j+1]}e_{j+1,j+1}$. Without
confusion, hereafter we adopt the following convention:\bea
e_{i,i'}&=&E^{i,i'}_{(0)},~~~i,i'=1,\ldots,n+m,\\
h_j&=&h^j_{(0)}=(-1)^{[j]}e_{j,j}-(-1)^{[j+1]}e_{j+1,j+1},
~j=1,\ldots,n+m-1,
\\
E_{i,i}&=&(E^{i,i})_{1\ldots N}=\sum_{k=1}^{N}E^{i,i}_{(k)},~i=1,\ldots,n+m,\label{Q-gen-1}\\
H_j&=&(-1)^{[j]}E_{j,j}-(-1)^{[j+1]}E_{j+1,j+1},~j=1,\ldots,n+m-1,\\
E_{j,j+1}&=&(E^{j,j+1})_{1\ldots
N}=\sum_{k=1}^NE^{j,j+1}_{(k)}q^{\sum_{i=k+1}^Nh^j_{(i)}},
~j=1,\ldots,n+m-1,\\
E_{j+1,j}&=& (E^{j+1,j})_{1 \ldots
N}=\sum_{k=1}^Nq^{-\sum_{i=1}^{k-1}h^j_{(i)}}E^{j+1,j}_{(k)},
~j=1,\ldots,n+m-1,\label{Q-gen-2}\eea and  similar convention for
the non-simple generators. Then the operators $\{E_{i,j}\}$ are
the operators which act {\it non-trivially\/} on the quantum space
${\cal{H}}$ and {\it trivially}  (i.e. as an identity)  on the
auxiliary space $V_0$. From (\ref{Sc}), we have \bea
\P^{\s_c}\,(h^j)_{0\ldots N}\,(\P^{\s_c})^{-1}&=&h_j+H_j,~j=1,\ldots,n+m-1,\label{t-eq-3}\\
\P^{\s_c}\,(E^{j,j+1})_{0\ldots
N}\,(\P^{\s_c})^{-1}&=&q^{h_j}\,E_{j,j+1}+e_{j,j+1},
~j=1,\ldots,n+m-1,\\
\P^{\s_c}\,(E^{j+1,j})_{0 \ldots
N}\,(\P^{\s_c})^{-1}&=&e_{j,j+1}q^{-H_j}+E_{j+1,j},
~j=1,\ldots,n+m-1. \label{t-eq-4} \eea Using
(\ref{t-eq-1})-(\ref{t-eq-2}), (\ref{t-eq-3})-(\ref{t-eq-4}) and
Lemma 1, we have

\vskip0.1in

\begin{Proposition}  \bea
&&(E^{\g,\g-l})_{0\ldots N}=q^{-\sum_{k=1}^lh_{\g-k}}E_{\g,\g-l}+e_{\g,\g-l}\no\\
&&\quad\quad\hspace{-0.1truecm}+\sum_{i=1}^{l-1}(1\hspace{-0.1truecm}-\hspace{-0.1truecm}
q^{2(-1)^{[\g-l+i]}})q^{-\sum_{k=1}^{l-i}h_{\g-k}}
e_{\g-l+i,\g-l}\,E_{\g,\g-l+i},~\g-l\geq 1~{\rm and}~l\geq
2,\\
&&\P^{\s_c}(E^{\g,\g-l})_{0\ldots N}(\P^{\s_c})^{-1}
=e_{\g,\g-l}q^{-\sum_{k=1}^lH_{\g-k}}+E_{\g,\g-l}\no\\
&&\quad\quad\hspace{-0.1truecm}+\sum_{i=1}^{l-1}(1\hspace{-0.1truecm}-\hspace{-0.1truecm}
q^{2(-1)^{[\g-l+i]}})\,e_{\g,\g-l+i}\,
q^{-\sum_{k=1}^{l-i}H_{\g-k}}E_{\g-l+i,\g-l},~\g-l\geq 1~{\rm
and}~l\geq
2,\\
&&(E^{\g,\g+l})_{0\ldots N}=E_{\g,\g+l}+e_{\g,\g+l}q^{\sum_{k=0}^{l-1}H_{\g+k}}\no\\
&&\quad\quad\hspace{-0.1truecm}+\sum_{i=1}^{l-1}(1\hspace{-0.1truecm}-\hspace{-0.1truecm}
q^{-2(-1)^{[\g+i]}})e_{\g+i,\g+l}\,E_{\g,\g+i}q^{\sum_{k=i}^{l-1}H_{\g+k}}
,~\g+l\leq n+m~{\rm and}~l\geq 2,\label{dec-e-1}\\
&&\P^{\s_c}\,(E^{\g,\g+l})_{0\ldots N}\,(\P^{\s_c})^{-1}
=e_{\g,\g+l}+q^{\sum_{k=0}^{l-1}h_{\g+k}}\,E_{\g,\g+l}\no\\
&&\quad\quad\hspace{-0.1truecm}+\sum_{i=1}^{l-1}(1\hspace{-0.1truecm}-\hspace{-0.1truecm}
q^{-2(-1)^{[\g+i]}})e_{\g,\g+i}\,q^{\sum_{k=i}^{l-1}h_{\g+k}}E_{\g+i,\g+l}
,~\g+l\leq n+m~{\rm and}~l\geq 2, \label{dec-e-2}\eea
\end{Proposition}
\vskip0.1in

\noindent Substituting (\ref{dec-e-1}) and (\ref{dec-e-2}) into
Proposition 1, we obtain our main result in this section:

\vskip0.1in
\begin{Theorem} \label{Theo-1}
The matrix elements $T_{n+m,n+m-l}(u)\,(l=1,\ldots,n+m-1)$ of the monodromy matrix
can be expressed  in terms of $T_{n+m,n+m}(u)$ and the generators
of $\GL$ by the following recursive relation: \bea
&&T_{n+m,n+m-l}(u)=\no\\
&&\quad=\lt(q^{-(-1)^{[n+m]}}E_{n+m-l,n+m}T_{n+m,n+m}(u)-T_{n+m,n+m}(u)E_{n+m-l,n+m}\rt)
q^{-\sum_{k=1}^{l}H_{n+m-k}}\no\\
&&\quad\quad-\sum_{\a=1}^{l-1}(1-q^{-2(-1)^{[n+m-\a]}})T_{n+m,n+m-\a}(u)E_{n+m-l,n+m-\a}
q^{-\sum_{k=\a+1}^{l}H_{n+m-k}}.\label{Recursive}\eea
\end{Theorem}

\vskip0.1in

\noindent The proof of this theorem is relegated to Appendix A.

\vskip0.1in

\noindent We call the second term in the  R.H.S. of
(\ref{Recursive}) {\it quantum correction term}, which vanishes in
the rational limit ($q\rightarrow 1$). Moreover, such a nontrivial
correction term only occurs in the higher rank models (i.e., when
$n+m\geq 3$). In the rational limit: $q\rightarrow 1$,
(\ref{Recursive}) reduces to the (anti)commutation relations used
in \cite{Albert00,ZYZ05}. For some special values of $m$ and $n$,
the associated recursive relations in the present grading become:
\begin{itemize}
\item For the $U_q(gl(1|1))$  case: \bea
T_{2,1}(u)=\lt[q^{-1}E_{1,2}T_{2,2}(u)-T_{2,2}(u)E_{1,2}\rt]q^{-H_1}.\eea

\item For the $\gl$ case which corresponds to the quantum $t-J$
model: \bea
&&T_{3,2}(u)=\lt[q^{-1}E_{2,3}T_{3,3}(u)-T_{3,3}(u)E_{2,3}\rt]q^{-H_2},\\
&&T_{3,1}(u)=\lt[q^{-1}E_{1,3}T_{3,3}(u)-T_{3,3}(u)E_{1,3}\rt]q^{-H_2-H_1}\no\\
&&\quad\quad\quad\quad\quad-(1-q^2)T_{3,2}(u)E_{1,2}q^{-H_1}.\eea

\item For the $U_q(gl(2|2))$ case which corresponds to the quantum
EKS model \cite{Ess92}:\bea
&&T_{4,3}(u)=\lt[q^{-1}E_{3,4}T_{4,4}(u)-T_{4,4}(u)E_{3,4}\rt]q^{-H_3},\\
&&T_{4,2}(u)=\lt[q^{-1}E_{2,4}T_{4,4}(u)-T_{4,4}(u)E_{2,4}\rt]q^{-H_3-H_2}\no\\
&&\quad\quad\quad\quad\quad-(1-q^{-2})T_{4,3}(u)E_{2,3}q^{-H_2},\\
&&T_{4,1}(u)=\lt[q^{-1}E_{1,4}T_{4,4}(u)-T_{4,4}(u)E_{1,4}\rt]q^{-H_3-H_2-H_1}\no\\
&&\quad\quad\quad\quad\quad-(1-q^{-2})T_{4,3}(u)E_{1,3}q^{-H_2-H_1}\no\\
&&\quad\quad\quad\quad\quad-(1-q^{2})T_{4,2}(u)E_{1,2}q^{-H_1}.
\eea
\end{itemize}

%%%%%%%%%%%%%%%%%%%%%%%%%%%%%%%%%%%%%%%%%%%%%%%%%%%%%%%%%%%%%%
%                                                            %
%     Factorizing  F-matrices and their inverses             %
%                                                            %
%                                                            %
%%%%%%%%%%%%%%%%%%%%%%%%%%%%%%%%%%%%%%%%%%%%%%%%%%%%%%%%%%%%%%

\section{Factorizing  F-matrices and their inverses}
\label{F-matrix} \setcounter{equation}{0} In this section, we
construct the Drinfeld twists \cite{Drinfeld83} (factorizing
F-matrices) on the $N$-fold tensor product space (i.e. the quantum
space ${\cal{H}}$) associated with the quantum superalgebra $\GL$.
\subsection{Factorizing F-matrix}
Let $\S_{N}$ be the permutation group associated with the indices
$(1,\ldots,N)$  and $R^{\s}_{1\ldots N}$ the $N$-site $R$-matrix
associated with $\s\in\S_N$. $R^{\s}_{1\ldots N}$ acts
non-trivially on the quantum space ${\cal{H}}$ and trivially  (i.e
as an identity) on the auxiliary space.

\vskip0.in

\begin{Definition} The F-matrix $F_{1\ldots N}(z_1,\ldots,z_N)$
is an operator in ${\rm End}({\cal{H}})$ and satisfies the
following three properties:
\begin{itemize}
\item I. lower-triangularity; \item II. non-degeneracy; \item III.
factorization, namely,\bea
F_{\s(1)\ldots\s(N)}(z_{\s(1)},\ldots,z_{\s(N)})\, R^{\s}_{1\ldots
N}=F_{1\ldots N}(z_1,\ldots,z_N),~\forall \s\in \S_N.\eea
\end{itemize}
\end{Definition}
\vskip0.1in

\noindent Define the $N$-site $F$-matrix:
\begin{eqnarray}
F_{1\ldots N}\equiv F_{1\ldots N}(z_1,\ldots,z_N)=\sum_{\sigma\in
{\cal S}_N}
   {\sum_{\alpha_{\sigma(1)}\ldots\alpha_{\sigma(N)}=1}^{n+m}}^{\hspace{-0.6truecm}*}
   \hspace{0.6truecm}\prod_{j=1}^N P_{\sigma(j)}^{\alpha_{\sigma(j)}}
   S(\sigma,\alpha_\sigma)R_{1\ldots N}^\sigma, \label{de:F}
\end{eqnarray}
where $P^{\a}_i$ is the embedding of the project operator $P^{\a}$
in the $i$-th space  with $(P^{\a})_{kl}=\d_{kl}\d_{k\a}$.
 %Here and below we have used the convention
 %$\prod_{i=1}^nf_i=f_1f_2\ldots f_n $.
The sum $\sum^*$ in
(\ref{de:F}) is over all non-decreasing sequences of the labels
$\alpha_{\sigma(i)}$:
\begin{eqnarray}
&& \alpha_{\sigma(i+1)}\geq \alpha_{\sigma(i)}\quad \mbox{if}\quad
              \sigma(i+1)>\sigma(i), \nonumber\\
&& \alpha_{\sigma(i+1)}> \alpha_{\sigma(i)}\quad \mbox{if}\quad
              \sigma(i+1)<\sigma(i), \label{cond:F}
\end{eqnarray}
and $S(\sigma,\alpha_\sigma)$ is a c-number function of $\s,\a_\s$
and the element $c_{ij}$ of the R-matrix, defined by
\begin{eqnarray}
S(\sigma,\alpha_\sigma)\equiv
\exp\lt\{\frac{1}{2}\sum_{l>k=1}^N\lt(1-(-1)^{[\a_{\s(k)}]}\rt)\,
\delta_{\alpha_{\sigma(k)},\alpha_{\sigma(l)}}
    \ln(1+c_{\sigma(k)\sigma(l)})\rt\}.\label{F-1}
\end{eqnarray}

\vskip0.1in

\begin{Proposition} The F-matrix $F_{1\ldots N}$ given by
(\ref{de:F})-(\ref{F-1}) satisfies the properties I, II, III.
\end{Proposition}

\vskip0.1in

\noindent {\it Proof}. The definition of $F_{1\ldots N}$
(\ref{de:F}) and the summation condition (\ref{cond:F}) imply that
$F_{1\ldots N}$ is a lower-triangular matrix. Moreover, one can
easily check that the $F$-matrix is non-degenerate because all
diagonal elements are non-zero.

We now prove that the $F$-matrix (\ref{de:F}) satisfies the
property III. Any given permutation $\sigma\in {\cal S}_N$ can be
decomposed into elementary ones of the group ${\cal S}_N$ as
$\sigma=\sigma_{i_1}\ldots \sigma_{i_k}$.  By (\ref{elemet-1}), we
have, if the property III holds for any elementary permutation
$\sigma_i$,
\begin{eqnarray}
&&F_{\sigma(1\ldots N)}R^{\sigma}_{1\ldots N}= \nonumber\\
 &=& F_{\sigma_{i_1}\ldots\sigma_{i_k}(1\ldots N)}
     R^{\sigma_{i_k}}_{\sigma_{i_1}\ldots\sigma_{i_{k-1}}(1\ldots N)}
     R^{\sigma_{i_{k-1}}}_{\sigma_{i_1}\ldots\sigma_{i_{k-2}}(1\ldots N)}
     \ldots
     R^{\sigma_{i_1}}_{1\ldots N}\nonumber\\
 &=&F_{\sigma_{i_1}\ldots\sigma_{i_{k-1}}(1\ldots N)}
     R^{\sigma_{i_{k-1}}}_{\sigma_{i_1}\ldots\sigma_{i_{k-2}}(1\ldots N)}
     \ldots
     R^{\sigma_{i_1}}_{1\ldots N}\nonumber\\
  &=&\ldots
     =F_{\sigma_{i_1}(1\ldots N)}R^{\sigma_{i_1}}_{1\ldots N}=F_{1\ldots N}.
\end{eqnarray}

For the elementary permutation $\sigma_i$, we have
\begin{eqnarray}
 &&F_{\sigma_i(1\ldots N)}R^{\sigma_i}_{1\ldots N}= \nonumber\\
 &=&\sum_{\sigma\in {\cal S}_N}
   \sum_{\alpha_{\sigma_i\sigma(1)}\ldots\alpha_{\sigma_i\sigma(N)}}^{\quad\quad *}
   \prod_{j=1}^N P_{\sigma_i\sigma(j)}^{\alpha_{\sigma_i\sigma(j)}}
 %  \nonumber\\ &&\times
  S(\sigma_i\sigma,\alpha_{\sigma_i\sigma})R_{\sigma_i(1\ldots N)}^\sigma
   R_{1\ldots N}^{\sigma_i} \nonumber\\
 &=&\sum_{\sigma\in {\cal S}_N}
   \sum_{\alpha_{\sigma_i\sigma(1)}\ldots\alpha_{\sigma_i\sigma(N)}}^{\quad\quad *}
   \prod_{j=1}^N P_{\sigma_i\sigma(j)}^{\alpha_{\sigma_i\sigma(j)}}
%   \nonumber\\ &&\times
   S(\sigma_i\sigma,\alpha_{\sigma_i\sigma})
   R_{1\ldots N}^{\sigma_i\sigma} \nonumber\\
 &=&\sum_{\tilde\sigma\in {\cal S}_N}
   \sum_{\alpha_{\tilde\sigma(1)}\ldots\alpha_{\tilde\sigma(N)}}^{\quad\quad *(i)}
   \prod_{j=1}^N P_{\tilde\sigma(j)}^{\alpha_{\tilde\sigma(j)}}
   S(\tilde\sigma,\alpha_{\tilde\sigma})R_{1\ldots N}^{\tilde\sigma}, \label{eq:FR-F}\nonumber\\
\end{eqnarray}
where $\tilde\sigma=\sigma_i\sigma$, and the summation sequences
of $\alpha_{\tilde\sigma}$ in ${\sum^*}^{(i)}$ now has the form
\begin{eqnarray}
&& \alpha_{\tilde\sigma(j+1)}\geq \alpha_{\tilde\sigma(j)}\quad
\mbox{if}\quad
              \sigma_i\tilde\sigma(j+1)>\sigma_i\tilde\sigma(j), \nonumber\\
&& \alpha_{\tilde\sigma(j+1)}> \alpha_{\tilde\sigma(j)}\quad
\mbox{if}\quad
              \sigma_i\tilde\sigma(j+1)<\sigma_i\tilde\sigma(j). \label{cond:FR-F}
\end{eqnarray}
Comparing (\ref{cond:FR-F}) with (\ref{cond:F}), we find that the
only difference between them is the transposition $\sigma_i$
factor in the ``if" conditions.  For a given $\tilde\sigma\in
{\cal S}_N$ with $\tilde\sigma(j)=i$ and $\tilde\sigma(k)=i+1$, we
now examine how the elementary transposition $\sigma_i$ will
affect the inequalities (\ref{cond:FR-F}). If $|j-k|>1$, then
$\sigma_i$ does not affect the sequence of $\alpha_{\tilde\sigma}$
at all, that is, the sign of inequality $``>"$ or ``$\geq$"
between two neighboring root indexes is unchanged with the action
of $\sigma_i$. If $|j-k|=1$, then in the summation sequences of
$\alpha_{\tilde\sigma}$, when $\tilde\sigma(j+1)=i+1$ and
$\tilde\sigma(j)=i$, sign ``$\geq$" changes to $``>"$, while when
$\tilde\sigma(j+1)=i$ and $\tilde\sigma(j)=i+1$, $``>"$ changes to
``$\geq$". Thus (\ref{cond:F}) and (\ref{eq:FR-F}) differ only
when equal labels $\alpha_{\tilde\sigma}$ appear. With the help of
the relation $c_{21}c_{12}=1$, one may prove that in this case the
product $F_{\sigma_i(1\ldots N)}R^{\sigma_i}_{1\ldots N}$ still
equals $F_{1\ldots N}$ (see \cite{zsy04} for a more detailed
proof). Thus, we obtain
\begin{eqnarray}
R_{1\ldots N}^\sigma(z_1,\ldots,z_N)
 =F^{-1}_{\sigma(1\ldots N)}(z_{\sigma(1)},\ldots,
   z_{\sigma(N)})F_{1\ldots N}(z_1,\ldots,z_N),
\end{eqnarray}
and the factorizing $F$-matrix $F_{1\ldots N}$ of $\GL$ is proved
to satisfy all three properties.
\begin{flushright}$\Box$~~~~~\end{flushright}

\noindent From the expression of the $F$-matrix, one knows that it
has an even grading, i.e., \bea [F_{1\ldots N}]=0.\eea

\subsection{Inverse of the F-matrix}
The non-degenerate property of the $F$-matrix implies that we can
find the inverse matrix $F^{-1}_{1\ldots N}$. To do so, we first
define
\begin{eqnarray}
F^*_{1\ldots N}&=&\sum_{\sigma\in {\cal S}_N}
   {\sum_{\alpha_{\sigma(1)}\ldots\alpha_{\sigma(N)}=1}^{n+m}}
   ^{\hspace{-0.6truecm}\vspace{-.4truecm} **}
   S(\sigma,\alpha_\sigma)R_{\sigma(1\ldots N)}^{\sigma^{-1}}
   \prod_{j=1}^N P_{\sigma(j)}^{\alpha_{\sigma(j)}},
    \label{de:F*}
\end{eqnarray}
where the sum $\sum^{**}$ is taken over all possible $\alpha_i$
which satisfies the following non-increasing constraints:
\begin{eqnarray}
&& \alpha_{\sigma(i+1)}\leq \alpha_{\sigma(i)}\quad \mbox{if}\quad
              \sigma(i+1)<\sigma(i), \nonumber\\
&& \alpha_{\sigma(i+1)}< \alpha_{\sigma(i)}\quad \mbox{if}\quad
              \sigma(i+1)>\sigma(i). \label{cond:F*}
\end{eqnarray}
%and $S^*(c,\sigma,\alpha_\sigma)$ is given by
%\begin{eqnarray}
%S^*(c,\sigma,\alpha_\sigma)=\exp\{\sum_{k>l=1}^N\delta^{[3]}_{\alpha_{\sigma(k)},\alpha_{\sigma(l)}}
%    \ln(1+c_{\sigma(k)\sigma(l)})\}
%\end{eqnarray}

\vskip0.1in
\begin{Proposition}
The inverse of the $F$-matrix is given by
\begin{equation}
F^{-1}_{1\ldots N}=F^*_{1\ldots
N}\prod_{i<j}\Delta_{ij}^{-1},\label{Prop-4}
\end{equation}
where
\begin{eqnarray}
[\Delta_{ij}]^{\beta_i\beta_j}_{\alpha_i\alpha_j} =
 \delta_{\alpha_i\beta_i}\delta_{\alpha_j\beta_j}\left\{
 \begin{array}{cl}\displaystyle
  {\sinh(z_i-z_j)\over\sinh(z_i-z_j+\eta)}&
   \mbox{if} \ \alpha_i>\alpha_j\\ \displaystyle
  {\sinh(z_j-z_i)\over\sinh(z_j-z_i+\eta)}&
   \mbox{if} \ \alpha_i<\alpha_j,\\
  1& \mbox{if}\ \alpha_i=\alpha_j=m+1,\ldots,n+m, \\ \displaystyle
  {-4\sinh^2(z_i-z_j)\cosh^2\eta\over
  \sinh(z_i-z_j+\eta)\sinh(z_i-z_j-\eta)}&
  \mbox{if}\ \alpha_i=\alpha_j=1,\ldots,m.
  \end{array}\right. \label{de:F-Inv}
\end{eqnarray}

\end{Proposition}
\vskip0.1in

\noindent {\it Proof}. We compute the product of $F_{1\ldots N}$
and $F^*_{1\ldots N}$. Substituting (\ref{de:F}) and (\ref{de:F*})
into the product, we have
\begin{eqnarray}
 F_{1\ldots N}F^*_{1\ldots N}% \nonumber\\
 &=&\sum_{\sigma\in {\cal S}_N}\sum_{\sigma'\in {\cal S}_N}
    \sum^{\quad\quad *}_{\alpha_{\sigma_1}\ldots\alpha_{\sigma_N}}
    \sum^{\quad\quad **}_{\beta_{\sigma'_1}\ldots\beta_{\sigma'_{N}}}
    S(\sigma,\alpha_\sigma)S(\sigma',\beta_{\sigma'})
    \nonumber\\ &&\times
    \prod_{i=1}^N P_{\sigma(i)}^{\alpha_{\sigma(i)}}
    R^{\sigma}_{1\ldots N}R^{{\sigma'}^{-1}}_{\sigma'(1\ldots N)}
    \prod_{i=1}^N P_{\sigma'(i)}^{\beta_{\sigma'(i)}} \nonumber\\
 &=&\sum_{\sigma\in {\cal S}_N}\sum_{\sigma'\in {\cal S}_N}
    \sum^{\quad\quad *}_{\alpha_{\sigma_1}\ldots\alpha_{\sigma_N}}
    \sum^{\quad\quad **}_{\beta_{\sigma'_1}\ldots\beta_{\sigma'_{N}}}
    S(\sigma,\alpha_\sigma)S(\sigma',\beta_{\sigma'})
     \nonumber\\ &&\times
    \prod_{i=1}^N P_{\sigma(i)}^{\alpha_{\sigma(i)}}
    R^{{\sigma'}^{-1}\sigma}_{\sigma'(1\ldots N)}
    \prod_{i=1}^N P_{\sigma'(i)}^{\beta_{\sigma'(i)}}. \label{eq:FF*-1}
\end{eqnarray}
To evaluate the R.H.S., we examine the matrix element of the
$R$-matrix
\begin{eqnarray}
\left(R^{{\sigma'}^{-1}\sigma}_{\sigma'(1\ldots N)}\right)
  ^{\alpha_{\sigma(N)}\ldots\alpha_{\sigma(1)}}
  _{\beta_{\sigma'(N)}\ldots\beta_{\sigma'(1)}}.
  \label{eq:R-index}
\end{eqnarray}
Note that the sequence $\{ \alpha_{\sigma}\}$ is non-decreasing
and $\{\beta_{\sigma'}\}$ is non-increasing. Thus the
non-vanishing condition of the matrix element (\ref{eq:R-index})
requires that $\alpha_{\sigma}$ and $\beta_{\sigma'}$ satisfy
\begin{eqnarray}
\beta_{\sigma'(N)}=\alpha_{\sigma(1)},\ldots,
\beta_{\sigma'(1)}=\alpha_{\sigma(N)}. \label{re:alpha-beta}
\end{eqnarray}
One can verify \cite{Albert00} that (\ref{re:alpha-beta}) is
fulfilled only if
\begin{eqnarray}
\sigma'(N)=\sigma(1),\ldots,\sigma'(1)=\sigma(N).
\label{re:sigma-sigma'}
\end{eqnarray}

Let $\bar\sigma$ be the maximal element of the ${\cal S}_N$ which
reverses the site labels
\begin{eqnarray}
\bar\sigma(1,\ldots,N)=(N,\ldots,1).
\end{eqnarray}
Then from (\ref{re:sigma-sigma'}), we have
\begin{eqnarray}
\sigma'=\sigma\bar\sigma. \label{eq:sigma'}
\end{eqnarray}
Substituting (\ref{re:alpha-beta}) and (\ref{eq:sigma'}) into
(\ref{eq:FF*-1}), we have
\begin{eqnarray}
 F_{1\ldots N}F^*_{1\ldots N}%= \nonumber\\
 &=&\sum_{\sigma\in {\cal S}_N}
    \sum^{\quad\quad *}_{\alpha_{\sigma_1}\ldots\alpha_{\sigma_N}}
    S(\sigma,\alpha_\sigma)S(\sigma,\alpha_\sigma)
    \prod_{i=1}^N P_{\sigma(i)}^{\alpha_{\sigma(i)}}
    R^{\bar\sigma}_{\sigma(N\ldots 1)}
    \prod_{i=1}^N P_{\sigma(i)}^{\alpha_{\sigma(i)}}\label{eq:FF*-2}.
      \nonumber\\
\end{eqnarray}
The decomposition of $R^{\bar\sigma}$ in terms of elementary
$R$-matrices is unique module the GYBE. One reduces from
(\ref{eq:FF*-2}) that $FF^*$ is a diagonal matrix:
\begin{eqnarray}
F_{1\ldots N}F^*_{1\ldots N}=\prod_{i<j}\Delta_{ij}.
\end{eqnarray}
Then (\ref{Prop-4}) is a simple consequence of the above equation.
~~~~~~$\Box$

%%%%%%%%%%%%%%%%%%%%%%%%%%%%%%%%%%%%%%%%%%%%%%%%%%%%%%%%%%%%%%%%%%%
%                                                                 %
%                                                                 %
%                  Monodromy matrix in the $F$-basis              %
%                                                                 %
%                                                                 %
%%%%%%%%%%%%%%%%%%%%%%%%%%%%%%%%%%%%%%%%%%%%%%%%%%%%%%%%%%%%%%%%%%%

\section{Monodromy matrix in the $F$-basis}
\label{F-B} \setcounter{equation}{0}

In the previous section, we have constructed  the $F$-matrix and
its inverse which act on the quantum space ${\cal{H}}$. The
non-degeneracy of the $F$-matrix means that its column vectors
also form a complete basis of ${\cal{H}}$, which is called the
$F$-basis. In this section, we study the generators of $\GL$ and
the elements of the monodromy matrix in the $F$-basis.

\subsection{$\GL$ generators in the $F$-basis}

The Cartan generators $\{E^{i,i}\}$ (or $\{H^j\}$) and the simple
generators $\{E^{j,j+1}\}$, $\{E^{j+1,j}\}$ of  $\GL$ are realized
on ${\cal{H}}$  by $\{E_{i,i}\}$ (or $\{H_j\}$), $\{E_{j,j+1}\}$
and $\{E_{j+1,j}\}$, respectively, as
(\ref{Q-gen-1})-(\ref{Q-gen-2}). The other non-simple generators
$\{E^{i,j}\}$ can be obtained from the simple ones by
(\ref{non-simple1}) and (\ref{non-simple2}), and denote their
realizations on ${\cal{H}}$ by $\{E_{i,j}\}$. Introduce the
generators in the $F$-basis:\bea \tilde{E}_{i,j}&=&F_{1\ldots
N}E_{i,j}F^{-1}_{1\ldots N},~i,j=1,\ldots,n+m. \label{E}\eea

\vskip0.1in
\begin{Theorem}  In the $F$-basis the Cartan and the simple generators of
$\GL$ are given by\bea
\tilde{E}_{i,i}&=&E_{i,i}=\sum_{k=1}^{N}E^{i,i}_{(k)},~i=1,\ldots,n+m,
\label{H-form-1}\\
\tilde{E}_{j,j+1}&=&\sum_{k=1}^{N}E^{j,j+1}_{(k)}\otimes_{\g\neq
k}G^{j,j+1}_{(\g)}(k,\g),~j=1,\ldots,n+m-1,\\
\tilde{E}_{j+1,j}&=&\sum_{k=1}^{N}E^{j+1,j}_{(k)}\otimes_{\g\neq
k}G^{j+1,j}_{(\g)}(k,\g), ~j=1,\ldots,n+m-1.\eea Here the diagonal
matrices $G^{\g,\g\pm 1}_{(j)}(i,j)$ are:\begin{itemize} \item For
$1<\g+1\leq m$, \bea
(G^{\g,\g+1}_{(j)}(i,j))_{kl}&=&\d_{kl}\lt\{\begin{array}{ll}
2e^{-\eta}\cosh\eta,&k=\g,\\(2a_{ij}\cosh\eta)^{-1}\,e^{\eta},
&k=\g+1,\\
1,&{\rm otherwise},\end{array}\rt.\\[4pt]
(G^{\g+1,\g}_{(j)}(i,j))_{kl}&=&\d_{kl}\lt\{\begin{array}{ll}
2e^{-\eta}\cosh\eta,&k=\g+1,\\(2a_{ji}\cosh\eta)^{-1}\,e^{\eta},
&k=\g,\\
1,&{\rm otherwise},\end{array}\rt. \eea

\item For $\g= m$, \bea
(G^{\g,\g+1}_{(j)}(i,j))_{kl}&=&\d_{kl}\lt\{\begin{array}{ll}
2e^{-\eta}\cosh\eta,&k=\g,\\e^{-\eta},
&k=\g+1,\\
1,&{\rm otherwise},\end{array}\rt.\\[4pt]
(G^{\g+1,\g}_{(j)}(i,j))_{kl}&=&\d_{kl}\lt\{\begin{array}{ll}
(2a_{ji}\cosh\eta)^{-1}e^{\eta},&k=\g,\\(a_{ji})^{-1}\,e^{\eta},
&k=\g+1,\\
1,&{\rm otherwise},\end{array}\rt. \eea

\item For $1+m\leq\g< n+m$, \bea
(G^{\g,\g+1}_{(j)}(i,j))_{kl}&=&\d_{kl}\lt\{\begin{array}{ll}
(a_{ij})^{-1}\,e^{\eta},&k=\g,\\e^{-\eta},
&k=\g+1,\\
1,&{\rm otherwise},\end{array}\rt.\\[4pt]
(G^{\g+1,\g}_{(j)}(i,j))_{kl}&=&\d_{kl}\lt\{\begin{array}{ll}
(a_{ji})^{-1}\,e^{-\eta},&k=\g+1,\\e^{\eta},
&k=\g,\\
1,&{\rm otherwise}.\end{array}\rt. \eea

\end{itemize}

\end{Theorem}
\vskip0.1in

\noindent {\it Proof}. Using (\ref{Q-gen-1})-(\ref{Q-gen-2}),
Proposition 1 and the composition law of R-matrices
(\ref{elemet-1}), one can prove the Theorem. Here, without losing
the generality, we give the proof for the generator
$\tilde{E}_{1,2}$ as an example.

>From the expressions of  $F_{1\ldots N}$ and its inverse, we have
\begin{eqnarray}
 \tilde E_{1,2}%\nonumber\\
 &=&\sum_{\sigma,\sigma'\in {\cal S}_N}
    \sum^{\quad\quad*}_{\alpha_{\sigma(1)}\ldots\alpha_{\sigma(N)}}
    \sum^{\quad\quad**}_{\beta_{\sigma'(1)}\ldots\beta_{\sigma'(N)}}
    S(\sigma,\alpha_\sigma)S(\sigma',\beta_{\sigma'})
    \nonumber\\ && \times
    \prod_{i=1}^N P_{\sigma(i)}^{\alpha_{\sigma(i)}}
    R^{\sigma}_{1\ldots N}(E^{1,2})_{1\ldots N}
    R^{{\sigma'}^{-1}}_{\sigma'(1\ldots N)}
    \prod_{i=1}^N
    P_{\sigma'(i)}^{\beta_{\sigma'(i)}}\prod_{i<j}\Delta_{ij}^{-1}
     \nonumber\\
 &=&\sum_{\sigma,\sigma'\in {\cal S}_N}
    \sum^{\quad\quad*}_{\alpha_{\sigma(1)}\ldots\alpha_{\sigma(N)}}
    \sum^{\quad\quad**}_{\beta_{\sigma'(1)}\ldots\beta_{\sigma'(N)}}
    S(\sigma,\alpha_\sigma)S(\sigma',\beta_{\sigma'})
    \nonumber\\ &&\times
    \prod_{i=1}^N P_{\sigma(i)}^{\alpha_{\sigma(i)}}
    [({\cal P}^\sigma_{1\ldots N}(E^{1,2})_{1,\dots,N}
     {\cal P}^{\sigma^{-1}}_{1\ldots N})]
    R^{{\sigma'}^{-1}\sigma}_{\sigma'(1\ldots N)}
    \prod_{i=1}^N P_{\sigma'(i)}^{\beta_{\sigma'(i)}}
    \prod_{i<j}\Delta_{ij}^{-1}
    \label{eq:E-1}     \\
 &=&\sum_{\sigma,\sigma'\in {\cal S}_N}
    \sum_{k=1}^N E^{1,2}_{(\sigma(l))}
    q^{\sum_{i=l+1}^{N}h_{(\sigma(i))}^1}
    \sum^{\quad\quad*}_{\alpha_{\sigma(1)}\ldots\alpha_{\sigma(N)}}
    \sum^{\quad\quad**}_{\beta_{\sigma'(1)}\ldots\beta_{\sigma'(N)}}
    S(\sigma,\alpha_\sigma)S(\sigma',\beta_{\sigma'}) \nonumber\\ && \times
    P_{\sigma(1)}^{\alpha_{\sigma(1)}=1}\ldots
    \left(P_{\sigma(l)=k}^{\alpha_{\sigma(l)}=1\rightarrow 2}\right)
    \ldots P_{\sigma(N)}^{\alpha_{\sigma(N)}}% \nonumber\\ && \times
    R^{{\sigma'}^{-1}\sigma}_{\sigma'(1\ldots N)}
    \prod_{i=1}^N
    P_{\sigma'(i)}^{\beta_{\sigma'(i)}}\prod_{i<j}\Delta_{ij}^{-1},
    \label{eq:E-2}% \nonumber\\
\end{eqnarray}
where in (\ref{eq:E-1}) we have used (\ref{F-def}) and $l$ stands
for indices between 1 and N such that $\s(l)=k$. Here and below,
$1\rightarrow 2$ in (\ref{eq:E-2}) means that $\a_{\s(l)}=1$ is
replaced by $\a_{\s(l)}=2$ in the site $\s(l)=k$. The element of
$R^{{\sigma'}^{-1}\sigma}_{\sigma'(1\ldots N)}$ between\\ $
P_{\sigma(1)}^{\alpha_{\sigma(1)}=1}\ldots
    \left(P_{\sigma(l)=k}^{\alpha_{\sigma(l)}=1\rightarrow 2}\right)
    \ldots P_{\sigma(N)}^{\alpha_{\sigma(N)}}$ and $
 P_{\sigma'(N)}^{\beta_{\sigma'(N)}}\ldots
   P_{\sigma'(1)}^{\beta_{\sigma'(1)}}$
is denoted as
 \begin{eqnarray}
 \left(R^{{\sigma'}^{-1}\sigma}_{\sigma'(1\ldots N)}\right)
 ^{\stackrel{\sigma(N)}{\alpha_{\sigma(N)}}\ldots
 \stackrel{\sigma(l)=k}{1\rightarrow 2}
  \ldots \stackrel{\sigma(1)}{1}}
 _{\beta_{\sigma'(N)}\ldots \beta_{\sigma'(1)}}. \label{eq:E-R-index}
 \end{eqnarray}
We call the sequence $\{\alpha_{\sigma(l)}\}$ {\bf normal} if it
is arranged according to the rules in (\ref{cond:F}), otherwise,
we call it {\bf abnormal}.

It is now convenient for us to discuss the non-vanishing condition
of the $R$-matrix element (\ref{eq:E-R-index}).  Comparing
(\ref{eq:E-R-index}) with (\ref{eq:R-index}), we find that the
difference between them lies in the $k$th site. Because the group
label in the $k$th space has been changed, the sequence
$\{\alpha_\sigma\}$ is now a abnormal sequence. However, it can be
permuted to the normal sequence by some permutation $\hat
\sigma_k$. Namely, $\alpha_{1\rightarrow 2}$ in the abnormal
sequence  can be moved to a suitable position by using the
permutation $\hat \sigma_k$ according to rules in (\ref{cond:F}).
(It is easy to verify that $\hat\sigma_k$ is unique by using
(\ref{cond:F}).) Thus, by procedure similar to that in the
previous section, we find that when
\begin{equation}
\sigma'=\hat\sigma_k\sigma\bar\sigma\quad \mbox{and}\quad
 \beta_{\sigma'(N)}=\alpha_{\sigma(1)},\ldots,
 \beta_{\sigma'(1)}=\alpha_{\sigma(N)},\label{eq:E-sigma}
\end{equation}
the $R$-matrix element (\ref{eq:E-R-index}) is non-vanishing.
% (We note that for
%a $\{P_{\sigma}^{\alpha_\sigma}\}$ sequence
%(\ref{eq:sqc-alpha}), we can find $m-l+1$
%$\{P_{\sigma'}^{\beta_\sigma'}\}$ sequences which ensure the
%non-vanishing of $R$-matrix according to the (\ref{cond:F}) and
%(\ref{cond:F*})).

Because the non-zero condition of the elementary $R$-matrix
element $R^{i'j'}_{ij}$ is $i+j=i'+j'$, the following $R$-matrix
elements
\begin{eqnarray}
 \left(R^{{\sigma'}^{-1}\sigma}_{\sigma'(1\ldots N)}\right)
 ^{\stackrel{\sigma(N)}{\alpha_{\sigma(N)}}\ldots
 \stackrel{\sigma(l)=k}{1} \ldots
 \stackrel{\sigma(p)}{1\rightarrow 2}
  \ldots \stackrel{\sigma(1)}{1}}
 _{\beta_{\sigma'(N)}\ldots \beta_{\sigma'(1)}}, \label{eq:E-R-nindex}
 \end{eqnarray}
with $1\leq p\leq l$ are also non-vanishing.

Therefore, (\ref{eq:E-2}) becomes
\begin{eqnarray}
 \tilde E_{1,2}%\nonumber\\
 &=&\sum_{\sigma\in {\cal S}_N}\sum_{k=1}^N
    \sum^{\quad\quad *}_{\alpha_{\sigma_1}\ldots\alpha_{\sigma_N}}
    S(\sigma,\alpha_\sigma)
    S(\hat\sigma_k\sigma,\alpha_{\hat\sigma_k\sigma})
       \nonumber\\ && \times
    \left[E^{1,2}_{(\sigma(l))}
    q^{\sum_{i=l+1}^{N}h_{(\sigma(i))}^1}
          P_{\sigma(1)}^{\alpha_{\sigma(1)}=1}
    \ldots P_{\sigma(l)=k}^{\alpha_{\sigma(l)}=1\rightarrow 2}
    \ldots P_{\sigma(N)}^{\alpha_{\sigma(N)}}\right.
      \nonumber\\ && \mbox{} \quad
          +\ldots \nonumber\\ && \mbox{} \quad
   +E^{1,2}_{(\sigma(p))}q^{\sum_{i=n+1}^{N}h_{(\sigma(i))}^1}
    P_{\sigma(1)}^{\alpha_{\sigma(1)}=1}
    \ldots P_{\sigma(p)}^{\alpha_{\sigma(p)}=1\rightarrow 2} \ldots
 %   \nonumber\\ &&\quad\quad\quad\times
    P_{\sigma(l)=k}^{\alpha_{\sigma(l)}=1}\ldots
    P_{\sigma(N)}^{\alpha_{\sigma(N)}}
     \nonumber\\ && \mbox{} \quad + \ldots
       \nonumber\\ && \mbox{}  \mbox{}\quad\left.
   +E^{1,2}_{(\sigma(1))}q^{\sum_{i=2}^{N}h_{(i)}^1}
    P_{\sigma(1)}^{\alpha_{\sigma(1)}=1\rightarrow 2}
    \ldots P_{\sigma(l)=k}^{\alpha_{\sigma(l)}=1}\ldots
    P_{\sigma(N)}^{\alpha_{\sigma(N)}}\right]
        \nonumber\\ && \times
    R^{\bar\sigma\sigma^{-1}\hat\sigma^{-1}_k\sigma}
     _{\hat\sigma_k\sigma(N\ldots 1)}
    \prod_{i=1}^N
    P_{\hat\sigma_k\sigma(i)}^{\alpha_{\hat\sigma_k\sigma(i)}}
    \prod_{i<j}\Delta_{ij}^{-1} \label{eq:E-3}\\
 &=&\sum_{k=1}^N E^{1,2}_{(k)}\otimes_{j\ne k} G^{1,2}_{(j)}(k,j),
     \label{eq:E-12-tilde}
\end{eqnarray}
where index $p$ runs between 1 and $l$ and $\hat\sigma_k$ is the
element of ${\cal S}_N$ which permutes the first abnormal sequence
in the square bracket of (\ref{eq:E-3}) to normal sequence. Using
the similar procedure, one can prove the Theorem for other
generators. ~~$\Box$

\vskip0.1in

\noindent The non-simple generators $\tilde{E}_{\g,\g\pm l}$ (for
$l\geq 2$) can be obtained through the simple ones by
(\ref{non-simple1}) and (\ref{non-simple2}).

Some remarks are in order. (\ref{H-form-1}) implies that \bea
\tilde{H}_j=H_j,~~j=1,\ldots,N,\eea which will be  used
frequently. In the rational limit: $\eta\rightarrow 0$ (or
$q\rightarrow 1$), our results reduce to  those in \cite{ZYZ05}
and those in \cite{Albert00} for the special case $m=0$.

\subsection{Creation operators in the
$F$-basis}

\noindent Among the matrix elements of the mondromy matrix
$T_{i,j}(u)$, the operators $T_{n+m,n+m-l}(u)$
($l=1,\ldots,n+m-1$) are called creation operators \cite{Kor93}
and  are usually denoted by \bea
C_{n+m-l}(u)=T_{n+m,n+m-l}(u),~~l=1,\ldots,n+m-1.\eea In the
$F$-basis, they become  \bea \tilde{C}_{n+m-l}(u)=F_{1\ldots
N}C_{n+m-l}(u)F^{-1}_{1\ldots N},~
l=1,\ldots,n+m-1.\label{Creation}\eea Let us denote
$T_{n+m,n+m}(u)$ by $D(u)$ and the corresponding operator in the
$F$-basis by\bea \tilde{D}(u)=F_{1\ldots N}D(u)F^{-1}_{1\ldots N}.
\label{D-operator}\eea

\vskip0.1in

\begin{Proposition} $\tilde{D}(u)$ is a diagonal
matrix given by
\begin{eqnarray}
\tilde D(u)=\otimes_{i=1}^N
  \mbox{diag}\left(a_{0i},\ldots,a_{0i},1\right)_{(i)}.
  \label{eq:T33-tilde}
\end{eqnarray}
\end{Proposition}

\vskip0.1in

\noindent {\it Proof}. From (\ref{def-T}), we derive that  \bea
D(u)P^{n+m}_{0}=T_{n+m,n+m}(u)e_{n+m,n+m}=P^{n+m}_{0}T_{0,1\ldots
N}(u)P^{n+m}_{0}.\eea Acting the $F$-matrix  from the left on the
both sides of the above equation, we have
\begin{eqnarray}
 F_{1\ldots N}D(u)P^{n+m}_0
 &=&\sum_{\sigma\in {\cal S}_N}
    \sum_{\alpha_{\sigma(1)}\ldots\alpha_{\sigma(N)}}^{\quad\quad*}
    S(\sigma,\alpha_\sigma)\prod_{i=1}^N
    P_{\sigma(i)}^{\alpha_{\sigma}}R^\sigma_{1\ldots N}
    P_0^{m+n} T_{0,1\ldots N}(u)P_0^{m+n} \nonumber\\
 &=&\sum_{\sigma\in {\cal S}_N}
    \sum_{\alpha_{\sigma(1)}\ldots\alpha_{\sigma(N)}}^{\quad\quad*}
    S(\sigma,\alpha_\sigma)\prod_{i=1}^N
    P_{\sigma(i)}^{\alpha_{\sigma}}
    P_0^{m+n} T_{0,\sigma(1\ldots N)}(u)P_0^{m+n}
    R^\sigma_{1\ldots N}.\no\\
\end{eqnarray}
Following \cite{Albert00}, we can split the sum $\sum^*$ according
to the number of occurrences of the index $m+n$.
\begin{eqnarray}
F_{1\ldots N}D(u)P^{n+m}_0
 &=&\sum_{\sigma\in {\cal S}_N}
    \sum_{k=0}^N
    \sum_{\alpha_{\sigma(1)}\ldots\alpha_{\sigma(N)}}^{\quad\quad*}
    S(\sigma,\alpha_\sigma)
    \prod_{j=N-k+1}^N \delta_{\alpha_{\sigma(j)},{m+n}}
    P_{\sigma(j)}^{\alpha_{\sigma(j)}} \nonumber\\ &&\times
    \prod_{j=1}^{N-k}P_{\sigma(j)}^{\alpha_{\sigma(j)}}
    P_0^{m+n} T_{0,\sigma(1\ldots N)}(u)P_0^{m+n}
    R^\sigma_{1\ldots N}. \label{eq:T-tilde-1}
\end{eqnarray}
Consider the prefactor of $R^\sigma_{1\ldots N}$. We have
\begin{eqnarray}
 &&\prod_{j=1}^{N-k}P_{\sigma(j)}^{\alpha_{\sigma(j)}}
   \prod_{j=N-k+1}^N
    P_{\sigma(j)}^{{m+n}}
    P_0^{m+n} T_{0,\sigma(1\ldots N)}(u)P_0^{m+n}\nonumber\\
 &=&\prod_{j=1}^{N-k}P_{\sigma(j)}^{\alpha_{\sigma(j)}}
   \prod_{j=N-k+1}^N\left(
   R_{0\,\sigma(j)}\right)^{{m+n}\ {m+n}}_{{m+n}\ {m+n}}
   P_0^{m+n} T_{0,\sigma(1\ldots N-k)}(u)P_0^{m+n}
   \prod_{j=N-k+1}^N P_{\sigma(j)}^{{m+n}} \nonumber\\
 &=&
   \prod_{j=1}^{N-k}P_{\sigma(j)}^{\alpha_{\sigma(j)}}
   P_0^{m+n} T_{0,\sigma(1\ldots N-k)}(u)P_0^{m+n}
   \prod_{j=N-k+1}^N P_{\sigma(j)}^{{m+n}} \nonumber\\
 &=&
   \prod_{i=1}^{N-k}\left(R_{0\,\sigma(i)}\right)
            ^{{m+n}\alpha_{\sigma(i)}}_{{m+n}\alpha_{\sigma(i)}}
   \prod_{j=1}^{N-k}P_{\sigma(j)}^{\alpha_{\sigma(j)}}
   \prod_{j=N-k+1}^N P_{\sigma(j)}^{{m+n}}\,\,P^{n+m}_0\nonumber\\
 &=&
   \prod_{i=1}^{N-k}a_{0\,\sigma(i)}
   \prod_{j=1}^{N-k}P_{\sigma(j)}^{\alpha_{\sigma(j)}}
   \prod_{j=N-k+1}^N P_{\sigma(j)}^{{m+n}}\,\,P^{n+m}_0,  \label{eq:T-tilde-2}
\end{eqnarray}
where $a_{0i}=a(u,z_i)$. Substituting (\ref{eq:T-tilde-2}) into
(\ref{eq:T-tilde-1}), we have
\begin{eqnarray}
F_{1\ldots N}D(u)=\otimes_{i=1}^N
  \mbox{diag}\left(a_{0i},\ldots,a_{0i},1\right)_{(i)}
 F_{1\ldots N}.
\end{eqnarray}
This completes the proof of the proposition. ~~~~~~$\Box$

\vskip 0.1in

By means of the expressions of the generators of $\GL$ in the
previous subsection, combining with  the Theorem 1 and Proposition
5, we have

\vskip 0.1in

\begin{Theorem}  In the $F$-basis the creation operators $C_{n+m-l}(u)$
$(l=1,\ldots,n+m-1)$ are given by \bea
\tilde{C}_{n+m-l}(u)&=&\lt(q^{-(-1)^{[n+m]}}\tilde{E}_{n+m-l,n+m}
\tilde{D}(u)-\tilde{D}(u)\tilde{E}_{n+m-l,n+m}\rt)
q^{-\sum_{k=1}^{l}H_{n+m-k}}\no\\
&&\hspace{-0.1truecm}-\hspace{-0.1truecm}
\sum_{\a=1}^{l-1}(1\hspace{-0.1truecm}-\hspace{-0.1truecm}
q^{-2(-1)^{[n+m-\a]}})\tilde{C}_{n+m-\a}(u)
\tilde{E}_{n+m-l,n+m-\a}
q^{-\sum_{k=\a+1}^{l}H_{n+m-k}}.\label{Recursive1}\eea
\end{Theorem}

\vskip0.1in

\noindent For some special values  of $m$ and $n$, we have:
\begin{itemize}

\item For $m=0$ and $n=2$ (namely, the $U_q(gl(2)$ case), our
general results reduce to  those in \cite{Maillet96}.

%\item For $U_q(gl(1|1))$ case: \bea \tilde C(u)&=&
%  (q^{-1}\tilde E_{1,2}\tilde D(u)-\tilde D(u)\tilde E_{1,2})
%  q^{-H_1}
% \nonumber\\
%  &=&\sum_{i=1}^N{e^{-(u-u_i)}\sinh\eta\over\sinh(u-u_i+\eta)}
%     E^{12}_{(i)}\otimes_{j\ne i}\mbox{diag}
%     \left({2\cosh\eta\sinh(u-u_j)\over
%            \sinh(u-u_j+\eta)},1\right)_{(j)}.\label{eq:C-gl11} \eea

 \item For the $\gl$ case,

 \begin{eqnarray}
\tilde C_2(u)&=&
 \left(q^{-1}\tilde E_{2,3}\tilde D(u)-\tilde D(u)\tilde E_{2,3}\right)
 q^{-H_2}\nonumber\\
 &=&\sum_{i=1}^N{e^{-(u-z_i)}\sinh\eta\over \sinh(u-z_i+\eta)}
 E^{2,3}_{(i)}\nonumber\\ &&
 \otimes_{j\ne i}\mbox{diag}\left(
 {\sinh(u-z_j)\over\sinh(u-z_j+\eta)},
 {2\sinh(u-z_j)\cosh\eta\over\sinh(u-z_j+\eta)},1\right)_{(j)},
 \label{eq:C2-tilde}\\
%%%%%%%%%%%%%%%%%%%%%%%%%%%%%%%%%%%%%%%%%%%%%%%%%%%%%%%%%%%%%%%%%%%%%%%%%
%
%%%%%%%%%%%%%%%%%%%%%%%%%%%%%%%%%%%%%%%%%%%%%%%%%%%%%%%%%%%%%%%%%%%%%%%%%
\tilde C_1(u)&=&
 \left(q^{-1}\tilde E_{1,3}\tilde D(u)-\tilde D(u)\tilde E_{1,3}\right)
 q^{-H_1-H_2}-(1-q^2)\tilde C_2(u)\tilde E_{1,2}q^{-H_1} \nonumber\\
 &=&\sum_{i=1}^N{e^{-(u-z_i)}\sinh\eta\over \sinh(u-z_i+\eta)}
 E^{1,3}_{(i)}
 \otimes_{j\ne i}\mbox{diag}\left(
 {2\sinh(u-z_j)\cosh\eta\over\sinh(u-z_j+\eta)},
 \right.\nonumber\\ &&\left.
 {\sinh(u-z_j)\sinh(z_i-z_j+\eta)\over
 \sinh(z_i-z_j)\sinh(u-z_j+\eta)},1\right)_{(j)}\nonumber\\
 && \mbox{}+\sum_{i\ne j=1}^N
  {e^{z_j-u}\sinh(u-z_i)\sinh^2\eta
   [e^{z_j-z_i}+2\sinh(z_i-z_j)]\over
  \sinh(u-z_i+\eta)\sinh(u-z_j+\eta)\sinh(z_i-z_j)}
  E^{1,2}_{(i)}\otimes E^{2,3}_{(j)}\nonumber\\
 && \otimes_{k\ne i,j}\left(
 {2\sinh(u-z_k)\cosh\eta\over\sinh(u-z_k+\eta)},
 %\right.\nonumber\\ &&\left.
 {\sinh(u-z_k)\sinh(z_i-z_k+\eta)\over
 \sinh(z_i-z_k)\sinh(u-z_k+\eta)},1\right)_{(k)}.
 \nonumber\\
 \label{eq:C1-tilde}
\end{eqnarray}
\end{itemize}

%%%%%%%%%%%%%%%%%%%%%%%%%%%%%%%%%%%%%%%%%%%%%%%%%%%%%%%%%%%%%%%%%
%                                                               %
%        Bethe vectors in the $F$-basis                         %
%                                                               %
%%%%%%%%%%%%%%%%%%%%%%%%%%%%%%%%%%%%%%%%%%%%%%%%%%%%%%%%%%%%%%%%%

\section{Bethe vectors in the $F$-basis}
\label{B-F} \setcounter{equation}{0}

The explicit expressions of the creation operators of the $\GL$
monodromy matrix in the $F$-basis, given in previous section,
enable us to resolve the hierarchy of the nested Bethe vectors of
the model associated with $\GL$. Here, we  take the quantum $t-J$
model (i.e. the  $\gl$-model) as an example to demonstrate the
procedure. The generalization to the general $\GL$ case  is
straightforward.

In the framework of the standard algebraic Bethe ansatz method
\cite{Fad79}, the Bethe vector of the quantum supersymmetric
$t$-$J$ model is given by
\begin{eqnarray}
\Omega_N=\sum_{d_1\ldots d_{\a}}
  (\Omega^{(1)}_{\a})^{d_1\ldots d_{\a}}C_{d_1}(v_1)\ldots
  C_{d_{\a}}(v_{\a})|vac\rangle, \label{de:Omega}
\end{eqnarray}
where $\a$ is a positive integer and  $d_i=1,2$, $|vac\rangle$ is
the pseudo-vacuum state
\begin{equation}
|vac\rangle=\otimes_{i=1}^N\left(\begin{array}{c}
0\\0\\1\end{array} \right)_{(i)}, \label{de:vacuum-tj}
\end{equation}
and $(\Omega^{(1)}_{\a})^{d_1\ldots d_{\a}}$ are functions of the
spectral parameters $v_j$, and are the vector components of the
nested Bethe vector $\Omega^{(1)}$. The nested Bethe vector
$\Omega^{(1)}$ is  given by
\begin{equation}
\Omega_{\a}^{(1)}=C^{(1)}(v^{(1)}_1) C^{(1)}(v^{(1)}_2)\cdots
          C^{(1)}(v^{(1)}_{\b})|vac\rangle^{(1)}, \label{de:Bethe-vector-nested}
\end{equation}
where $\b$ is a positive integer and  $|vac\rangle^{(1)}$ is the
nested pseudo-vacuum state
\begin{equation}
|vac\rangle^{(1)}=\otimes_{i=1}^{\a}\left(\begin{array}{c}
0\\1\end{array} \right)_{(i)}. \label{de:vacuum-nested}
\end{equation}
The creation operator of the nested $U_q(gl(2))$ system
$C^{(1)}(u)$ is the lower-triangular entry of the nested monodromy
matrix $T^{(1)}(v^{(1)})$
\begin{eqnarray}
T^{(1)}(v^{(1)})&=&r_{0\a}(v^{(1)}-v_{\a})r_{0\,\a-1}(v^{(1)}-v_{\a-1})
 \ldots r_{01}(v^{(1)}-v_1)
\nonumber\\
&\equiv& \left(\begin{array}{ccc}
 A^{(1)}(v^{(1)})& B^{(1)}(v^{(1)})\\
 C^{(1)}(v^{(1)})&
 D^{(1)}(v^{(1)})\end{array}\right),\label{Nested-Mon}
\end{eqnarray}
and the nested $R$-matrix is
\begin{equation}
r_{12}(u_1,u_2)\equiv r_{12}(u_1-u_2)= \left(
 \begin{array}{cccc}
 c_{12}&0&0&0\\ 0&a_{12}&-b^+_{12}&0\\
 0&-b^-_{12}&a_{12}&0\\ 0&0&0&c_{12} \end{array}\right).
 \label{de:r-nested}
\end{equation}
%One easily sees that $\Omega^{(1)}$ spans a subspace of the space
%spanned by $\Omega_N$ through (\ref{de:Omega}).
(\ref{Nested-Mon}) implies that the nested system is a
$U_q(gl(2))$ spin chain on $\a$-site lattice and the corresponding
inhomogeneous parameters are $\{v_{i}|i=1,\ldots,\a\}$. So, in the
following, we shall adopt the same convention for the the nested
system as that in previous sections but the inhomogeneous
parameters will be  replaced by $\{v_i\}$.

Acting the associated $F$-matrix on the pseudo-vacuum state
(\ref{de:vacuum-tj}), one finds that the pseudo-vacuum state is
invariant. It is due to the fact that only terms with roots equal
to 3 will produce  non-zero results. Therefore, the $U_q(gl(2|1))$
Bethe vector (\ref{de:Omega}) in the $F$-basis can be written as
\begin{eqnarray}
   \tilde\Omega_N(v_1,\ldots,v_{\a})
 &\equiv& F_{1\ldots N}\Omega_N(v_1,\ldots,v_{\a})\nonumber\\
 &=& \sum_{d_1\ldots d_{\a}}
  (\Omega^{(1)}_{\a})^{d_1\ldots d_{\a}}\tilde C_{d_1}(v_1)\ldots \tilde
  C_{d_{\a}}(v_{\a})|vac\rangle. \label{eq:phi-F}
\end{eqnarray}
The $c$-number coefficient $(\Omega^{(1)}_{\a})^{d_1\ldots
d_{\a}}$ has to be evaluated in the original basis, not in the
$F$-basis.

\subsection{Nested Bethe vectors in the $F$-basis}
Let us first compute the nested Bethe vectors in the $F$-basis.
For the nested $R$-matrix (\ref{de:r-nested}), we define the
$\a$-site $F$ and $F^*$-matrices by
\begin{eqnarray}
F^{(1)}_{1\ldots \a}&=&\sum_{\sigma\in {\cal S}_{\a}}
   {\sum_{\alpha_{\sigma(1)}\ldots\alpha_{\sigma(\a)}=1}^2}^
   {\vspace{-0.4truecm}\hspace{-0.6truecm}*}~~~~~
   \prod_{j=1}^{\a} P_{\sigma(j)}^{\alpha_{\sigma(j)}}
   \bar S^{(1)}(c,\sigma,\alpha_\sigma)r_{1\ldots \a}^\sigma,
   \label{de:F-nested} \\
F^{*(1)}_{1\ldots \a}&=&\sum_{\sigma\in {\cal S}_{\a}}{
   \sum_{\alpha_{\sigma(1)}\ldots\alpha_{\sigma(\a)}=1}^2}
   ^{\vspace{-0.4truecm}\hspace{-0.6truecm}**}~~~~~
\bar S^{(1)}(c,\sigma,\alpha_\sigma)r_{\sigma(1\ldots
\a)}^{\sigma^{-1}}
   \prod_{j=1}^{\a} P_{\sigma(j)}^{\alpha_{\sigma(j)}},
    \label{de:F*-nested}
\end{eqnarray}
respectively, where the $c$-number function $\bar S^{(1)}$ is
given by
\begin{eqnarray}
\bar S^{(1)}(c,\sigma,\alpha_\sigma)\equiv
\exp\{\sum_{l>k=1}^{\a}\delta_{\alpha_{\sigma(k)},\alpha_{\sigma(l)}}
    \ln(1+c_{\sigma(k)\sigma(l)})\}.
\end{eqnarray}
Therefore the inverse of the $F$-matrix can be represented in
terms of  the $F^*$-matrix as
\begin{eqnarray}
(F^{(1)}_{1\ldots \a})^{-1}=F^{*(1)}_{1\ldots \a}\prod_{i<j}
 (\bar \Delta^{(1)}_{ij})^{-1}, \label{eq:F-inverse-nested}
\end{eqnarray}
with
\begin{eqnarray*}
\bar \Delta^{(1)}_{ij}&=&\mbox{diag}\left(
{4\sinh^2(v_i-v_j)\cosh^2\eta\over
  \sinh(v_i-v_j+\eta)\sinh(v_i-v_j-\eta)}
,{\sin(v_j-v_i)\over\sinh(v_j-v_i+\eta)},
\right. \nonumber\\
&&\quad\quad \quad\left. {\sinh(v_i-v_j)\over\sinh(v_i-v_j+\eta)},
{4\sinh^2(v_i-v_j)\cosh^2\eta\over
  \sinh(v_i-v_j+\eta)\sinh(v_i-v_j-\eta)} \right).\nonumber\\
\end{eqnarray*}

With the help of the $F$-matrix (\ref{de:F-nested}) and its
inverse, one may compute the $\a$-site $U_q(gl(2))$ generator
$\tilde E_{1,2}$ in the $F$-basis,
\begin{eqnarray}
\tilde E_{1,2}=\sum_{i=1}^{\a}E^{1,2}_{(i)}\otimes_{j\ne i}\left(
2e^{-\eta}\cosh\eta,
                 {e^{\eta}\sinh(v_i-v_j+\eta)\over
                  2\sinh(v_i-v_j)\cosh\eta}\right)_{(j)}.
 \label{eq:E-12-nested}
\end{eqnarray}
Define $\tilde D^{(1)}(u)=F_{1\ldots
\a}^{(1)}D^{(1)}(u)(F^{(1)}_{1\ldots\a})^{-1}$. From Proposition
5, we obtain
\begin{equation}
\tilde D^{(1)}(u)=\otimes_{i=1}^{\a} \left(
 {\sinh(u-v_i)\over\sinh(u-v_i+\eta)},
 {\sinh(u-v_i-\eta)\over\sinh(u-v_i+\eta)} \right)_{(i)}.
 \label{eq:D-nested}
\end{equation}
The  creation operator in the $F$-basis is then obtained with the
help of  the nested $R$-matrix (\ref{de:r-nested}) and Theorem 3
in the case of  $m=2$ and $n=0$, and is given by
\begin{eqnarray}
\tilde C^{(1)}(u)&=&
  (q\tilde E_{1,2}\tilde D(u)-\tilde D(u)\tilde E_{1,2})
  q^{-H_1}\nonumber\\
  &=&-\sum_{i=1}^{\a}
  {e^{-(u-v_i)}\sinh\eta\over \sinh(u-v_i+\eta)}
  E^{1,2}_{(i)}\otimes_{j\ne i}\left(
  {2\sinh(u-v_j)\cosh\eta\over\sinh(u-v_j+\eta)},
  \right. \nonumber\\ && \quad\quad \left.
  {\sinh(v_i-v_j+\eta)\sinh(u-v_j-\eta)\over
   2\sinh(v_i-v_j)\sinh(u-v_j+\eta)\cosh\eta}
  \right)_{(j)},\label{eq:C-nested}
\end{eqnarray}

Applying $F^{(1)}_{1\dots\a}$ to the nested Bethe vector
$\Omega^{(1)}_{\a}$ (\ref{de:Bethe-vector-nested}), we obtain
\begin{eqnarray}
  \tilde\Omega^{(1)}_{\a}(v_1^{(1)},\ldots, v_{\b}^{(1)})
 &\equiv& F^{(1)}_{1\ldots \a}
          \Omega^{(1)}(v_1^{(1)},\ldots, v_{\b}^{(1)}) \nonumber\\
 &=&s(c)\tilde C^{(1)}(v^{(1)}_1)
\tilde C^{(1)}(v^{(1)}_2)\ldots
           \tilde C^{(1)}(v^{(1)}_{\b})|vac\rangle^{(1)},
\label{eq:Bethe-vector-nested-tilde}
\end{eqnarray}
where we have used
$F^{(1)}_{1\ldots\a}|vac\rangle^{(1)}=\prod_{i<j}(1+c_{ij})|vac\rangle^{(1)}\equiv
s(c)|vac\rangle^{(1)}$.

Substituting  ${\tilde C}^{(1)}(v)$ into
(\ref{eq:Bethe-vector-nested-tilde}), we obtain
\begin{eqnarray}
 &&\tilde \Omega^{(1)}_{\a}(v_1^{(1)},\ldots,v_{\b}^{(1)})
   =s(c)\tilde{C}^{(1)}(v_1^{(1)})\ldots \tilde{C}^{(1)}(v_{\b}^{(1)})
   \,|vac\rangle^{(1)}\nonumber\\
&=&s(c)\sum_{i_1<\ldots< i_{\b}}
B_{\b}^{(1)}(v_1^{(1)},\ldots,v_{\b}^{(1)}|v_{i_1},\ldots,v_{i_{\b}})
E_{(i_1)}^{1,2}\ldots E_{(i_{\b})}^{1,2}\,|vac\rangle^{(1)}\;,
\label{Psi_2}
\end{eqnarray}
where
\begin{eqnarray}
&& B^{(1)}_{\b}(v_1^{(1)},\ldots,v_{\b}^{(1)}|v_{1},\ldots,v_{\b})
\nonumber\\
 &=&\sum_{\sigma\in {\cal S}_{\b}}\prod_{k=1}^{\b}\left(-
 {e^{-(v_k^{(1)}-v_{\sigma(k)})}\sinh\eta\over
  \sinh(v_k^{(1)}-v_{\sigma(k)}+\eta)}\right)
\nonumber\\ && \times
  \prod^{\a}_{j\ne \sigma(k),\ldots,\sigma(\b)}
  {\sinh(v_{\sigma(k)}-v_j+\eta)\sinh(v_k^{(1)}-v_j-\eta)\over
   2\sinh(v_{\sigma(k)}-v_j)\sinh(v_k^{(1)}-v_j+\eta)\cosh\eta}
   \nonumber\\ &&\times
 \prod_{l=k+1}^{\b}
 {2\cosh\eta\sinh(v_k^{(1)}-v_{\sigma(l)})\over
  \sinh(v_k^{(1)}-v_{\sigma(l)}+\eta)}. \label{eq:B1}
\end{eqnarray}

\subsection{ Bethe vectors of the quantum supersymmetric $t$-$J$
model  in the $F$-basis} Now back to the Bethe vector
(\ref{eq:phi-F}) of the quantum supersymmetric $t$-$J$  model. As
is shown in Appendix B,  the Bethe vector is invariant (module
overall factor) under the exchange of arbitrary spectral
parameters:
\begin{eqnarray}
\tilde\Omega_N(v_{\sigma(1)},\ldots,v_{\sigma(\a)})
 ={1\over c^\sigma_{1\ldots \a}}
  \tilde\Omega_N(v_1,\ldots,v_{\a}),~\s\in \S_{\a}, \label{eq:exchange}
\end{eqnarray}
where $c^\sigma_{1\ldots \a}$ has the decomposition law
\begin{eqnarray}
 c^{\sigma'\sigma}_{1\ldots \a}
 =c^{\sigma}_{\sigma'(1\ldots \a)}
  c^{\sigma'}_{1\ldots \a}\label{eq:c-cc},
\end{eqnarray}
and  $c^{\sigma_i}_{1\ldots \a}=c_{i\ i+1}\equiv c(v_i,v_{i+1})$
for an elementary permutation $\sigma_i$.

This result is a generalization of that in \cite{Tak83,Vega89}.
This invariance  enables us to concentrate on a particularly
simple term in the sum (\ref{eq:phi-F}) of the following form with
$p_1$ number of $d_i=1$ and $\a-p_1$ number of $d_j=2$
\begin{eqnarray}
 \tilde C_1(v_1)\ldots\tilde C_1(v_{p_1})
 \tilde C_2(v_{p_1+1})\ldots\tilde C_2(v_{\a}).
% \equiv g_{1\ldots 12\ldots 2}(v_1,\ldots,p_1,p_{1}+1,\ldots n).
     \label{eq:C1-C2}
\end{eqnarray}
In the $F$-basis, the commutation relation between $C_i(v)$ and
$C_j(u)$, i.e. (\ref{eq:commu-cc-b}), becomes
\begin{eqnarray}
 \tilde C_i(v)\tilde C_j(u)
 &=&-{1\over a(u,v)}\tilde C_j(u)\tilde C_i(v)
    +{b(u,v)\over a(u,v)}\tilde C_j(v)\tilde C_i(u).
    \label{eq:commu-CC-F}
\end{eqnarray}
Then using (\ref{eq:commu-CC-F}), all $\tilde C_1$'s in
(\ref{eq:C1-C2}) can be moved to the right of all $\tilde C_2$'s,
yielding
\begin{eqnarray}
 &&\tilde C_1(v_1)\ldots\tilde C_1(v_{p_1})
   \tilde C_2(v_{p_1+1})\ldots\tilde C_2(v_{\a})= \nonumber\\
 &=&g(v_1,\ldots,v_{\a})
 \tilde C_2(v_{p_1+1})\ldots\tilde C_2(v_{\a})
 \tilde C_1(v_{1})\ldots\tilde C_1(v_{p_1})+\ldots\ ,
 \label{eq:C2-C1}
\end{eqnarray}
where $g(v_1,\ldots,v_{\a})=\prod_{k=1}^{p_1}\prod_{l=p_1+1}^{\a}
(-{1/ a(v_l,v_k)})$ is the contribution from the first term of
(\ref{eq:commu-CC-F}) and ``$\ldots$" stands for other terms
contributed by the second term of (\ref{eq:commu-CC-F}). It is
easy to see that these other terms have the form
\begin{equation}
\tilde C_2(v_{\sigma(p_1+1)})\ldots\tilde
C_2(v_{\sigma(\a)})\tilde C_1(v_{\sigma(1)})\ldots\tilde
C_1(v_{\sigma(p_1)}), \label{eq:C2-C1-sigma}
\end{equation}
with $\sigma\in {\cal S}_{\a}$. Substituting (\ref{eq:C2-C1}) into
the Bethe vector (\ref{eq:phi-F}), we obtain
\begin{eqnarray}
\tilde\Omega_N^{p_1}(v_1,\ldots,v_{\a})%= \nonumber\\
 &=&(\Omega^{(1)}_{\a})^{11\ldots 12\ldots 2}
    \prod_{k=1}^{p_1}~\prod_{l=p_1+1}^{\a}
    \left(-{1
      \over a(v_{l},v_{k})}\right) \nonumber\\
&&\times\;{\tilde C}_{2}(v_{p_1 +1})\ldots {\tilde C}_{2}(v_{\a})
{\tilde C}_{1}(v_{1})\ldots {\tilde C}_{1}(v_{p_1})|vac\rangle
+\ldots\, ,
 \label{eq:Phi-1122}
\end{eqnarray}
where and below, we  use the up-index $p_1$ to denote the Bethe
vector corresponding to the quantum number $p_1$. All other terms
in (\ref{eq:Phi-1122}) (denoted as ``\ldots") are to be obtained
from the first term by the permutation (exchange) symmetry. Then
we have,
\begin{eqnarray}
 &&{\tilde\Omega}_N^{p_1}(v_1,\ldots,v_{\a})= \nonumber\\
 &=&{1\over p_1!(\a-p_1)!}\sum_{\sigma \in {\cal S}_{\a}}
     c^{\sigma}_{1\ldots \a}
    (\Omega^{(1),\sigma}_{\a})^{11\ldots 12\ldots 2}
    \prod_{k=1}^{p_1}~\prod_{l=p_1+1}^{\a}
    \left(-{1\over a(v_{\sigma(l)},v_{\sigma(k)})}\right)
    \nonumber\\
&&\times\;{\tilde C}_{2}(v_{\sigma(p_1 +1)})\ldots {\tilde
C}_{2}(v_{\sigma(\a)}) {\tilde C}_{1}(v_{\sigma(1)})\ldots {\tilde
C}_{1}(v_{\sigma(p_1)})\,|vac\rangle\, , \label{eq:Phi-phi}
\end{eqnarray}
where $(\Omega^{(1),\sigma}_{\a})^{11\ldots 12\ldots 2}\equiv
(\hat f_\sigma\Omega^{(1)}_{\a})^{11\ldots 12\ldots 2}$ with $\hat
f_\sigma$ defined by (\ref{de:f-hat}) in the Appendix B.

We now show that $(\Omega^{(1)}_{\a})^{1\ldots 12\ldots 2}$ in
(\ref{eq:Phi-phi}), which has to be evaluated in the original
basis, is invariant (module an overall factor) under the action of
the $U_q(gl(2))$ $F$-matrix, i.e.
\begin{equation}
 (\Omega^{(1)}_{\a})^{11\ldots 12\ldots 2}
 =(t(c))^{-1}(\tilde\Omega^{(1)}_{\a})^{11\ldots 12\ldots 2},
\end{equation}
where the scalar factor $t(c)$ is \bea
t(c)=\prod_{j>i=1}^{p_1}(1+\bar{c}_{ij})
    \prod_{j>i=p_1+1}^{\a}(1+\bar{c}_{ij}),~\bar{c}_{ij}=c(v_i,v_j), \label{t-s}\eea
so that it can be expressed in the form of (\ref{Psi_2}).

Write the nested pseudo-vacuum vector in (\ref{de:vacuum-nested})
as
\begin{equation}
|vac\rangle^{(1)}\equiv |2\cdots 2\rangle^{(1)},
\end{equation}
where the number of 2 is $\a$. Then the nested Bethe vector
(\ref{eq:Bethe-vector-nested-tilde}) can be rewritten as
\begin{equation}
\Omega^{(1)}_{\a}(v_1^{(1)}\ldots v_{p_1}^{(1)})
 \equiv|\Omega^{(1)}_{\a}\rangle
 =\sum_{d_1\ldots d_{\a}}(\Omega^{(1)}_{\a})^{d_1\ldots d_{\a}}|d_1\ldots d_{\a}
 \rangle^{(1)}.
 \label{eq:phi1-phi1}
\end{equation}
Acting the $U_q(gl(2))$ $F$-matrix $F^{(1)}_{1\ldots\a}$ from left
on the above equation, we have
\begin{equation}
\tilde\Omega^{(1)}_{\a}(v_1^{(1)}\ldots v_{p_1}^{(1)})
 \equiv|\tilde\Omega^{(1)}_{\a}\rangle=F^{(1)}_{1\ldots\a}|\Omega^{(1)}_{\a}\rangle
 =\sum_{d_1\ldots d_{\a}}(\tilde\Omega^{(1)}_{\a})^{d_1\ldots d_{\a}}|d_1\ldots d_{\a}
 \rangle^{(1)}.
 \label{eq:phi1-phi1-F}
\end{equation}
It follows that
\begin{eqnarray}
  (\tilde\Omega^{(1)}_{\a})^{1\ldots 12\ldots2}
 &=&\langle1\ldots 12\ldots2|\tilde\Omega^{(1)}_{\a}\rangle
  =\langle1\ldots 12\ldots2|F_{1\ldots\a}^{(1)}|\Omega^{(1)}_{\a}
  \rangle\nonumber\\
 &=&\langle1\ldots 12\ldots2|\sum_{\sigma\in{\cal S}_{\a}}
    \sum^{\quad\quad*}_{\alpha_{\sigma(1)}\ldots\alpha_{\sigma(\a)}}
    \prod_{j=1}^{\a} P_{\sigma(j)}^{\alpha_{\sigma(j)}}
    \bar S^{(1)}(c,\sigma,\alpha_\sigma)
    R^\sigma_{1\ldots \a}
    |\Omega^{(1)}_{\a}\rangle \label{eq:tildephi-phi-1}\nonumber\\ \\
 &=&\langle1\ldots 12\ldots2|\left.\left\{
    \sum^{\quad\quad*}_{\alpha_{\sigma(1)}\ldots\alpha_{\sigma(\a)}}\prod_{j=1}^{\a}
    P_{\sigma(j)}^{\alpha_{\sigma(j)}}\right\}
    \right|_{\sigma=id}
    \bar S^{(1)}(c,\sigma,\alpha_\sigma)
    %R^{\sigma=id}_{1\ldots n}
    |\Omega^{(1)}_{\a}\rangle \label{eq:tildephi-phi-2} \nonumber\\ \\
 &=&t(c) \langle 1\ldots 12\ldots2|\Omega^{(1)}_{\a}\rangle %\nonumber\\
 =t(c)  (\Omega^{(1)}_{\a})^{1\ldots 12\ldots2},
 \label{eq:tildephi-phi-3}
\end{eqnarray}
where the scalar factor $t(c)$ is given by (\ref{t-s}).

Summarizing, we propose the following form of the $\gl$ Bethe
vector
\begin{eqnarray}
 &&{\tilde\Omega}_N^{p_1}(v_1,\ldots,v_{\a})= \nonumber\\
 &=&{1\over p_1!(\a-p_1)!}\sum_{\sigma \in {\cal S}_{\a}}
    c^{\sigma}_{1\ldots \a}%\left(
    \prod_{i=1}^{p_1}\prod_{j=p_1+1}^\a
    {2\sinh(v_{\sigma(i)}-v_{\sigma(j)})\cosh\eta\over
     \sinh(v_{\sigma(i)}-v_{\sigma(j)}+\eta)}
    %\right)%^{-1}
    \nonumber\\ && \times
    B^{(1)}_{p_1}(v_{1}^{(1)},\ldots,v_{p_1}^{(1)}|
    v_{\sigma(1)},\ldots,v_{\sigma(p_1)}) \nonumber\\
&&\times\;
    \prod_{k=1}^{p_1}\prod_{l=p_1+1}^{\a}
    \left(-{1\over a(v_{\sigma(l)},v_{\sigma(k)})}\right)
    {\tilde C}_{2}(v_{\sigma(p_1 +1)})\ldots {\tilde C}_{2}(v_{\sigma(\a)})
\nonumber\\ &&\times\;
    {\tilde C}_{1}(v_{\sigma(1)})\ldots {\tilde
C}_{1}(v_{\sigma(p_1)})\,|vac\rangle\, .   \label{eq:Phi-3a}
\end{eqnarray}

Substituting (\ref{eq:C2-tilde}) and (\ref{eq:C1-tilde}) into the
above relation, we finally have

\vskip0.1in
\begin{Proposition} The nested Bethe vector of the quantum $t-J$
model is given by
\begin{eqnarray}
&&{\tilde\Omega}_N^{p_1}(v_1,\ldots,v_{\a})\nonumber\\
 &=&{1\over p_1!(\a-p_1)!}
 \sum_{i_1<\ldots<i_{p_1}}\sum_{i_{p_1+1}<\ldots<i_{\alpha}}
  B_{\a,p_1}(v_1,\ldots,v_{\a};v_{1}^{(1)},\ldots,
              v_{p_1}^{(1)}|z_{i_1},\ldots,z_{i_{\a}})\nonumber\\
&&\times  \prod_{j={p_1+1}}^{{\a}}
E^{2,3}_{(i_j)}\prod_{j={1}}^{{p_1}}E^{1,3}_{(i_j)}\,|vac\rangle,
\end{eqnarray}
where
$\{i_1,i_2,\ldots,i_{p_1}\}\cap\{i_{p_1+1},i_{p_1+2},\ldots,i_{\alpha}\}=\varnothing$
and
\begin{eqnarray}
 &&B_{\a,p_1}(v_1,\ldots,v_{\a};v_{1}^{(1)},\ldots,
              v_{p_1}^{(1)}|z_{i_{1}},\ldots,z_{i_{\a}})=\nonumber\\
 &=&\sum_{\sigma \in S_{\a}}
    c^{\sigma}_{1\ldots \a}%\left(
    \prod_{i=1}^{p_1}~~
    \prod_{j=p_1+1}^{\a}
    {2\sinh(v_{\sigma(i)}-v_{\sigma(j)})\cosh\eta\over
     \sinh(v_{\sigma(i)}-v_{\sigma(j)}+\eta)}
    %\right)^{-1}
    \nonumber\\ && \times
  \prod_{k=1}^{p_1}~~\prod_{l=p_1+1}^{\a}
    \left(-{\sinh(v_{\sigma(l)}-z_{i_k})
           \sinh(v_{\sigma(l)}-v_{\sigma(k)}+\eta) \over
           \sinh(v_{\sigma(l)}-v_{\sigma(k)})
           \sinh(v_{\sigma(l)}-z_{i_k}+\eta)}\right)
      \nonumber\\ && \times
         B_{\a-p_1}^*(v_{\sigma(p_1+1)},\ldots,
                      v_{\sigma(\a)}|z_{i_{p_1+1}},\ldots,z_{i_{\a}})
                       \nonumber\\
&&\times
B_{p_1}^{(1)}(v_{1}^{(1)},\ldots,v_{p_1}^{(1)}|v_{\sigma(1)},\ldots,
                       v_{\sigma(p_1)})
         B_{p_1}^*(v_{\sigma(1)},\ldots,
                       v_{\sigma(p_1)}|z_{i_{1}},\ldots,z_{i_{p_1}}).\nonumber\\
\label{B1}
\end{eqnarray}
Here the function $B^*_p(v_1,\ldots,v_p|z_{1},\ldots,z_{p})$ is
given
\begin{eqnarray}
 && B^*_p(v_1,\ldots,v_p|z_{1},\ldots,z_{p})=\nonumber\\
 &=& \sum_{\sigma\in {\cal S}_p} \mbox{sign}(\sigma)
 \prod_{k=1}^p
 {e^{-(v_k-z_{\sigma(k)})}\sinh\eta\over
  \sinh(v_k-z_{\sigma(k)}+\eta)}
\prod_{l=k+1}^{p}
 {2\sinh(v_k-z_{\sigma(l)})\cosh\eta\over
   \sinh(v_k-z_{\sigma(l)}+\eta)}.\nonumber\\
\label{eq:B*}
\end{eqnarray}
\end{Proposition}

%%%%%%%%%%%%%%%%%%%%%%%%%%%%%%%%%%%%%%%%%%%%%%%%%%%%%%%%%%
%                                                        %
%                   Discussions                          %
%                                                        %
%%%%%%%%%%%%%%%%%%%%%%%%%%%%%%%%%%%%%%%%%%%%%%%%%%%%%%%%%%

\section{Discussions}
\label{Dis} \setcounter{equation}{0}

We have constructed the factorizing $F$-matrices for the
supersymmetric model associated with quantum superalgebra $\GL$
with generic $m$ and $n$, which includes the quantum
supersymmetric $t$-$J$ model as a special case. We have obtained
the completely symmetric representations for the creation
operators of the model in the $F$-basis. Our results make possible
a complete resolution of  the hierarchy of its nested Bethe
vectors. As an example, we have given  the explicit expressions of
the Bethe vectors of the quantum $t-J$ model in $F$-basis. Our
results are new even for the special $m=0$ case, which give the
results of the general model associated with $U_q(gl(n))$. (For
$m=0$ and $n=2$, our results reduce to those in \cite{Maillet96}.)

Authors in \cite{Korepin99} solved the quantum inverse problem of
the supersymmetric $t$-$J$ model in the original basis. Namely,
they reconstructed the local operators ($E^{i,j}_{(k)}$) in terms
of operators figuring in the $gl(2|1)$ monodromy matrix. Their
results should be  generalizable to the $\gl$ case. Then  together
with the results of the present paper in the $F$-basis one  should
be able to get the exact representations of form factors and
correlation functions of the quantum supersymmetric $t$-$J$ model.
These are under investigation and results will be reported elsewhere.\\[5mm]

%%%%%%%%%%%%%%%%%%%%%%%%%%%%%%%%%%%%%%%%%%%%%%%%%%%%%%%%%%%%%%%
%                                                             %
%  Acknowledgments                                            %
%                                                             %
%%%%%%%%%%%%%%%%%%%%%%%%%%%%%%%%%%%%%%%%%%%%%%%%%%%%%%%%%%%%%%%
\section*{Acknowledgements}
This work was financially supported by the Australia Research
Council. S.\,-Y. Zhao has also been supported by the UQ
Postdoctoral Research Fellowship.

\section*{Appendix A: Proof of Theorem \ref{Theo-1}}
\setcounter{equation}{0}
\renewcommand{\theequation}{A.\arabic{equation}}

\noindent Proposition 2 allows one to derive the following
equations:\bea
&&(E^{n+m-l,n+m})_{0\ldots N}=E_{n+m-l,n+m}+e_{n+m-l,n+m}q^{\sum_{k=1}^{l}H_{n+m-k}}\no\\
&&\quad\quad\hspace{-0.1truecm}+\sum_{\a=1}^{l-1}(1\hspace{-0.1truecm}-\hspace{-0.1truecm}
q^{-2(-1)^{[n+m-\a]}})e_{n+m-\a,n+m}\,E_{n+m-l,n+m-\a}q^{\sum_{k=1}^{\a}H_{n+m-k}}
,\no\\
&&\quad\quad\quad\quad l=1,\ldots,n+m-1,\label{dec-e-3}\\
&&\P^{\s_c}\,(E^{n+m-l,n+m})_{0\ldots N}\,(\P^{\s_c})^{-1}
=e_{n+m-l,n+m}+q^{\sum_{k=1}^{l}h_{n+m-k}}\,E_{n+m-l,n+m}\no\\
&&\quad\quad\hspace{-0.1truecm}+\sum_{i=1}^{l-1}(1\hspace{-0.1truecm}-\hspace{-0.1truecm}
q^{-2(-1)^{[n+m-\a]}})e_{n+m-l,n+m-\a}\,q^{\sum_{k=1}^{\a}h_{n+m-k}}E_{n+m-\a,n+m}
,\no\\
&&\quad\quad\quad l=1,\dots,n+m-1. \label{dec-e-4}\eea Taking
$x=E^{n+m-l,n+m}$ and $\s=\s_c$ and using (\ref{Mon}), then
(\ref{F-def}) becomes \bea T(u)\,(E^{n+m-l,n+m})_{0\ldots N}=
\P^{\s_c}\,(E^{n+m-l,n+m})_{0\ldots N}\,(\P^{\s_c})^{-1}
\,T(u).\eea Substituting (\ref{def-T}), (\ref{dec-e-3}) and
(\ref{dec-e-4}) into the above equation, we have, for the L.H.S.
of the resulting relation, \bea {\rm
L.H.S.}&=&\sum_{i=1}^{n+m}(-1)^{([i]+[n+m-l])([n+m-l]+[n+m]+1)}e_{i,n+m}T_{i,n+m-l}(u)
q^{\sum_{k=1}^lH_{n+m-k}}\no\\
&&\quad +\sum_{i,j=1}^{n+m}e_{i,j}(-1)^{[i]+[j]}T_{i,j}(u)E_{n+m-l,n+m}\no\\
&&\quad
+\sum_{i=1}^{n+m}\sum_{\a=1}^{l-1}(1-q^{-2(-1)^{[n+m-\a]}})(-1)^{([i]+[n+m-\a])([n+m-\a]+[n+m]+1)}\no\\
&&\quad\quad\times e_{i,n+m}T_{i,n+m-\a}(u)E_{n+m-l,n+m-\a}
q^{\sum_{k=1}^{\a}H_{n+m-k}}.\label{A-1} \eea Similarly for the
R.H.S. of the resulting relation, we obtain \bea {\rm
R.H.S.}&=&\sum_{i,j=1}^{n+m}(-1)^{([n+m-l]+[n+m]+1)([i]+[j])}
q^{\sum_{k=1}^{l}h_{n+m-k}}e_{i,j}E_{n+m-l,n+m}T_{i,j}(u)\no\\
&&\quad
+\sum_{i=1}^{n+m}(-1)^{[n+m]+[i]}e_{n+m-l,i}T_{n+m,i}(u)\no\\
&&\quad+\sum_{i=1}^{n+m}\sum_{\a=1}^{l-1}(1-q^{-2(-1)^{[n+m-\a]}})
(-1)^{([n+m-\a]+[i])([n+m-\a]+[n+m]+1)}\no\\
&&\quad\quad\times
e_{n+m-l,n+m-\a}q^{\sum_{k=1}^{\a}h_{n+m-k}}e_{n+m-\a,i}
E_{n+m-\a,n+m}T_{n+m-\a,i}(u).\label{A-2}\eea Comparing the
coefficients of the $e_{n+m,n+m}$ term on the both sides, we
obtain\bea
&&q^{-(-1)^{[n+m]}}\,E_{n+m-l,n+m}\,T_{n+m,n+m}(u)=\no\\
&&\quad\quad=
T_{n+m,n+m}(u)E_{n+m-l,n+m}+T_{n+m,n+m-l}(u)q^{\sum_{k=1}^{l}H_{n+m-k}}\no\\
&&\quad\quad\quad\quad+\sum_{\a=1}^{l-1}(1-q^{-2(-1)^{[n+m-\a]}})
T_{n+m,n+m-\a}(u)E_{n+m-l,n+m-\a}q^{\sum_{k=1}^{\a}H_{n+m-k}},
\eea which leads to the recursive relation (\ref{Recursive}).

\section*{Appendix B:$\quad$ The exchange symmetry of the Bethe vector}
\setcounter{equation}{0}
\renewcommand{\theequation}{B.\arabic{equation}}

For the Bethe vector $\Omega_N(v_1,\ldots,v_{\a})$ of the quantum
supersymmetric $t$-$J$ model, we define the exchange operator
$\hat f_\sigma=\hat f_{\sigma_{i_1}}\ldots \hat f_{\sigma_{i_k}}$
by
\begin{eqnarray}
 \hat f_\sigma \Omega_N(v_1,v_2,\ldots,v_{\a})
 =\Omega_N(v_{\sigma(1)},v_{\sigma(2)},\ldots,v_{\sigma(\a)}),
  \label{de:f-hat}
\end{eqnarray}
where $\sigma\in {\cal S}_{\a}$, and $\{\sigma_i\}$ are the
elementary permutations of ${\cal S}_{\a}$.

We first study the exchange symmetry for the elementary exchange
operator $\hat f_{\sigma_i}$ which exchanges the parameter $v_i$
and $v_{i+1}$. Acting $\hat f_{\sigma_i}$ on the Bethe vector of
$\gl$ (\ref{eq:phi-F}), we have
\begin{eqnarray}
&&\hat f_{\sigma_i} \Omega_N(v_1,v_2,\ldots,v_{\a})%\nonumber\\
 =\Omega_N(v_1,\ldots,v_{i+1},v_i,\ldots,v_{\a})\nonumber\\
 &=&\sum_{d_1,\ldots,d_{\a}}
 (\Omega^{(1),\sigma_i}_{\a})^{d_1\ldots d_{\a}} C_{d_1}(v _1)\ldots C_{d_i}(v_{i+1})
 C_{d_{i+1}}(v_i)\ldots C_{d_{\a}}(v _{\a})|vac\rangle, \label{eq:f-Phi}
\end{eqnarray}
where $\{(\Omega^{(1),\sigma_i}_{\a})^{d_1\ldots d_{\a}}\}$ are
the vector components of the nested Bethe vector
$\Omega^{(1),\sigma_i}_{\a}$ constructed by the nested monodromy
matrix
\begin{eqnarray}
T^{(1),\sigma_i}(u)&=&L_{\a}^{(1)}(u,v_{\a})\ldots
L_{i+1}^{(1)}(u,v_{i})
  L_{i}^{(1)}(u,v_{i+1})\ldots L_1^{(1)}(u,v_1),
\label{eq:mono-nest-i}
\end{eqnarray}
where the local $L$-operator is defined by
$L^{(1)}_i(u,v)=r_{0i}(u,v)$.

>From the GYBE (\ref{eq:GYBE-1}), one can derive the commutation
relation between $C_i(u)$ and $C_j(v)$, which is given by
\begin{eqnarray}
C_i(u)C_j(v)&=&\sum_{k,l}\check r(u,v)^{kl}_{ij} C_k(v)C_l(u).
\label{eq:commu-cc-b}
\end{eqnarray}
Here the braided $r$-matrix $\check{r}(u,v)\equiv {\cal P}r(u,v)$,
${\cal P}$ permutes the tensor product spaces of the 2-dimensional
$U_q(gl(2))$-module. Then, by (\ref{eq:commu-cc-b}),
(\ref{eq:f-Phi}) becomes
\begin{eqnarray}
\hat f_{\sigma_i} \Omega_N(v_1,v_2,\ldots,v_{\a})%\nonumber\\
&=&\sum_{d_1,\ldots,d_{\a}}
 (\Omega^{(1),\sigma_i}_{\a})^{d_1\ldots d_{\a}} C_{d_1}(v_1)\ldots
 \nonumber\\ &&\times
 (\check{r}(v_{i+1},v_i))^{k\ \ l}_{d_id_{i+1}}
 C_{k}(v_i)C_{l}(v_{i+1})\ldots
 C_{d_{\a}}(v_{\a})|vac\rangle.\label{eq:f-Phi-r}
\end{eqnarray}
We now compute the action of  $(\check r(v_{i+1},v_i))^{k\ \
l}_{d_id_{i+1}}$ on $(\Omega^{(1),\sigma_i})^{d_1\ldots d_{\a}}$.
One checks that $\check r$-matrix satisfies the YBE
\begin{eqnarray}
&&\check r_{i\,i+1}(v_{i+1},v_{i})
  L_{i+1}^{(1)}(u,v_{i})L_{i}^{(1)}(u,v_{i+1})\nonumber\\
&&\hspace{2em}=L_{i+1}^{(1)}(u,v_{i+1})L_{i}^{(1)}(u,v_{i})
  \check{r}_{i\,i+1}(v_{i+1},v_{i})\,.
\end{eqnarray}
Therefore, acting $\check r$ on $T^{(1),\sigma_i}(u)$, we have
\begin{eqnarray}
 \check{r}_{i\,i+1}(v_{i+1},v_{i})T^{(1),\sigma_i}(u)
 =T^{(1)}(u)\check{r}_{i\,i+1}(v_{i+1},v_{i}) .
\end{eqnarray}
Thus, because
$$\check r_{i\,i+1}(v_{i+1},v_{i})\, v_2\otimes v_2
 =c_{i\,i+1}(v_{i+1},v_{i})v_2\otimes v_2
 ={1\over c_{i\,i+1}}v_2\otimes v_2,$$
 we obtain
\begin{eqnarray}
 \sum_{d_i d_{i+1}}(\check r(v_{i+1},v_i))_{d_id_{i+1}}^{k\,\,\,l}
 (\Omega^{(1),\sigma_i}_{\a})^{d_1\ldots d_i d_{i+1}\ldots  d_{\a}}
 ={1\over c_{i\,i+1}}(\Omega^{(1)}_{\a})^{d_1\ldots kl\ldots d_{\a}}.
\end{eqnarray}
Changing the indices $k, l$ to $d_i,d_{i+1}$, respectively, and
substituting the above relation into (\ref{eq:f-Phi-r}), we obtain
the exchange symmetric relation of the Bethe vector of $\gl$
\begin{eqnarray}
\hat f_{\sigma_i} \Omega_N(v_1,v_2,\ldots,v_{\a})%\nonumber\\
&=&{1\over c_{i\,i+1}}\Omega_N(v_1,v_2,\ldots,v_{\a}),
\end{eqnarray}
for the elementary permutation operator $\sigma_i$.

It follows that under the action of the exchange operator
$f_{\sigma}$
\begin{eqnarray}
  \hat f_{\sigma} \Omega_N(v_1,v_2,\ldots,v_{\a})%\nonumber\\
 &=&{1\over c^\sigma_{1\ldots \a}}
  \Omega_N(v_1,v_2,\ldots,v_{\a}),
\end{eqnarray}
where $c^\sigma_{1\ldots \a}$ is defined in (\ref{eq:c-cc}).

%%%%%%%%%%%%%%%%%%%%%%%%%%%%%%%%%%%%%%%%%%%%%%%%%%%%%%%%%%%%%%%
%                                                             %
%  References                                                 %
%                                                             %
%%%%%%%%%%%%%%%%%%%%%%%%%%%%%%%%%%%%%%%%%%%%%%%%%%%%%%%%%%%%%%%

\end{document}